\documentclass[prd,twocolumn,aps,showpacs,groupedaddress,superscriptaddress]{revtex4}
\usepackage{graphicx,epsf,bm,bbm,color,amsmath,ulem,amssymb}
\usepackage[colorlinks=true,citecolor=green,linkcolor=red,urlcolor=blue]{hyperref}

    \newcommand{\tr}[1]{\textrm{#1}}

	\newcommand{\beq}{\begin{eqnarray}}

	\newcommand{\eeq}{\end{eqnarray}}
	\newcommand{\nn}{\nonumber}
	\newcommand{\q}{{\bf q}}
	
	\renewcommand{\k}{{\bf k}}
	\newcommand{\R}{{\bf R}}
    
	\newcommand{\K}{{\bf K}}

	\newcommand{\G}{{\bf G}}

\newcommand{\bem}{\begin{pmatrix}}
\newcommand{\eem}{\end{pmatrix}}

\newcommand{\f}{\frac}

\begin{document}
	
\title{Dirac points emerging from flat bands in Lieb-kagom\'e lattices}
\author{Lih-King Lim}
\email{lihking@zju.edu.cn}
\address{Zhejiang  Institute  of  Modern  Physics,  Department of Physics, Zhejiang  University,  Hangzhou, Zhejiang  310027, People's Republic of China}
\author{Jean-No\"el Fuchs}
\address{Sorbonne Universit\'e, CNRS, Laboratoire de Physique Th\' eorique de la Mati\` ere Condens\' ee, LPTMC, F-75005 Paris, France}     
\author{Fr\'ed\'eric Pi\'echon}
\address{Laboratoire de Physique des Solides, CNRS, Universit\'e Paris-Sud, Universit\'e Paris-Saclay, F-91405 Orsay, France}
\author{Gilles Montambaux}
\address{Laboratoire de Physique des Solides, CNRS, Universit\'e Paris-Sud, Universit\'e Paris-Saclay, F-91405 Orsay, France}

       \date{\today}
			
			\begin{abstract}
The energy spectra for the tight-binding models on the Lieb and kagom\'e lattices both exhibit a flat band. We study a model which continuously interpolates between these two limits. The flat band located in the middle of the three-band spectrum for the Lieb lattice is distorted, generating two pairs of Dirac points. While the upper pair evolves into graphene-like Dirac cones in the kagom\'e limit, the low energy pair evolves until it merges producing the band-bottom flat band. The topological characterization of the Dirac points is achieved by projecting the Hamiltonian on the two relevant bands in order to obtain an effective Dirac Hamiltonian. The low energy pair of Dirac points is particularly interesting: when they emerge, they have opposite winding numbers, but as they merge, they have the same winding number. This apparent paradox is due to a continuous rotation of their states in pseudo-spin space, characterized by a winding vector. This simple, but quite rich model, suggests a way to a systematic characterization of two-band contact points in multiband systems.
\end{abstract}

	\maketitle
\section{Introduction}
Multiband systems with flat band are of current interest thanks to their experimental realizations in a variety of systems \cite{Jo12,Guzman14,Taie15,Vicencio15,Muk14,Baboux16,Slot17}. The list includes cold atoms with optical lattices \cite{Jo12, Taie15}, photonic \cite{Guzman14,Vicencio15,Muk14} and polaritonic \cite{Baboux16} systems. Typically, the systems are engineered to faithfully realize exotic tight-binding models with multiple orbitals or atoms per unit cell, as well as complex lattice geometries. In these artificial systems, direct imaging of localized states \cite{Vicencio15, Slot17, Milicevic19} or the study of tunable interaction-induced effects \cite{Ozawa17} are few examples which show approaches hitherto not easily realizable in conventional condensed matter systems. 

\begin{figure}
\begin{center}
\includegraphics[width=8.8cm]{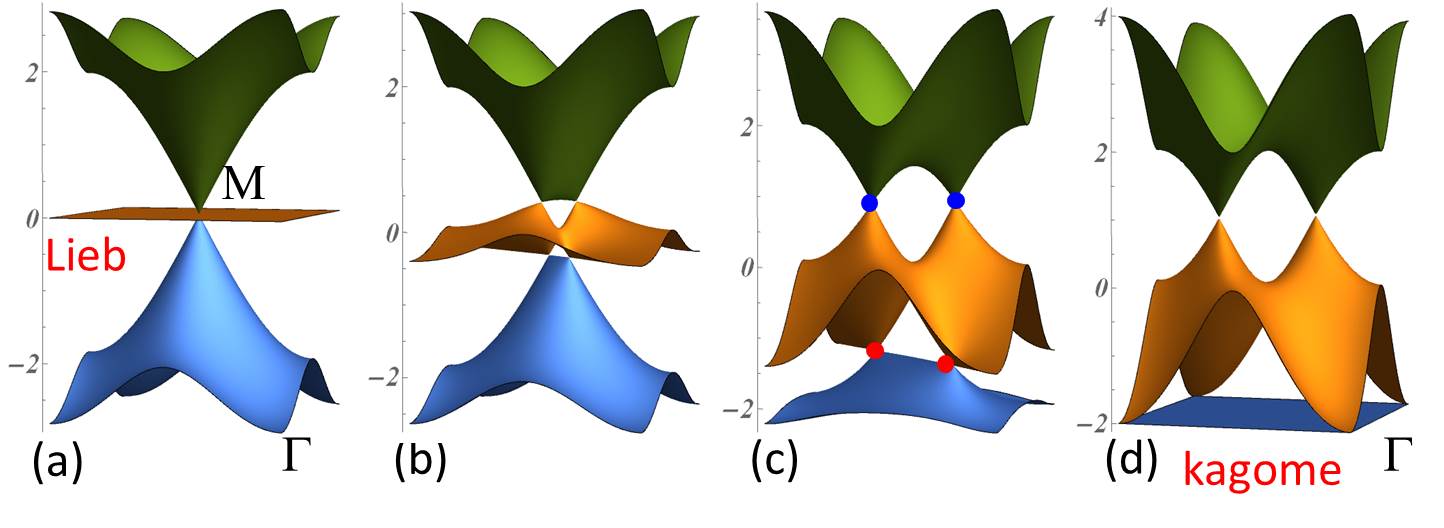}
\end{center}
\caption{Energy band spectrum of the Lieb-kagom\'e model in the Brillouin zone centered on the M point (see Fig. \ref{fig:BZ}a). (a) Lieb model ($t=0$); (b) $t=0.2$; (c) $t=0.7$; (d) kagom\'e model ($t=1$) in a square version. One pair of Dirac points is respectively formed in the upper two (blue dots) and lower two bands (red dots) away from the two limits.}
\label{fig:spectrum}
\end{figure}

\begin{figure}
\begin{center}
\includegraphics[width=8.5cm]{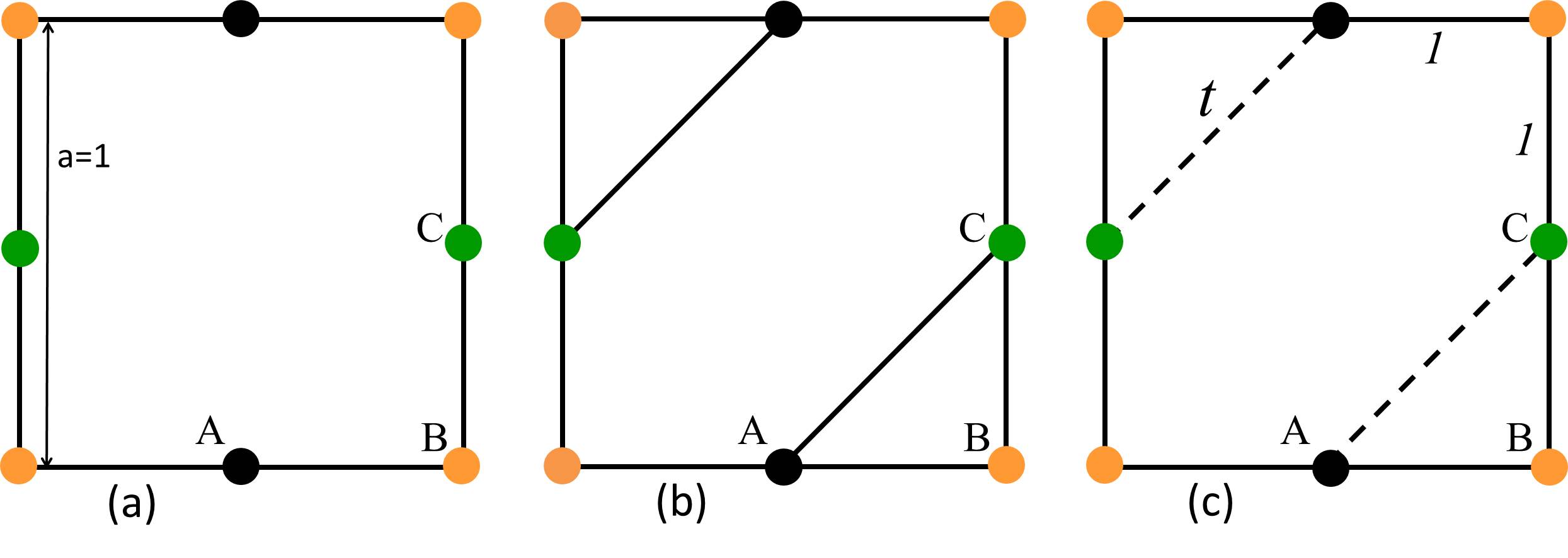}
\end{center}
 \caption{Unit cell of (a) the Lieb lattice (full lines represent unit hopping amplitudes); (b) the kagom\'e lattice in a square version; (c) the interpolating Lieb-kagom\'e lattice (dashed lines correspond to hopping amplitudes $0\leq t\leq 1$). The three sites in the unit cell are A,B and C.}
 \label{fig:lattice}
\end{figure}

In this work, we study a connection between two two-dimensional tight-binding models - on the Lieb lattice and on the kagom\'e lattice (also known as the trihexagonal tiling) - both featuring two dispersive bands and a flat band. On the one hand, the Lieb model~\cite{Dagotto1986,Apaja2010,Goldman2011,Tsai2015} has a band contact between three bands among which a topological flat band~\cite{Aoki1996} at zero energy (see Fig.~\ref{fig:spectrum}a, the tight-binding model is defined in Fig.~\ref{fig:lattice}a). On the other hand, the kagom\'e model~\cite{Ohgushi00,Xiao2003}, chosen here with a square symmetry, features a quadratic band contact point~\cite{Chong2008,Sun2009,Dora} between its lower bands: a dispersive band and a flat band resulting from destructive interferences~\cite{Aoki1996} (see Fig.~\ref{fig:spectrum}d). The real-space model is defined in Fig.~\ref{fig:lattice}b. Apart from the flat band, the kagom\'e energy spectrum is very similar to that of graphene with a pair of Dirac points between the upper bands (see Fig.~\ref{fig:spectrum}d). Both models have found physical realizations in the aforementioned systems \cite{Jo12,Guzman14,Taie15,Vicencio15,Muk14,Baboux16,Slot17}. There are also solid state systems expected to behave as a kagom\'e lattice for itinerant electrons, see e.g. Refs.~\cite{Mazin2014,Ye2018, Leykam2018}.

By interpolating between the two lattice models, we define the Lieb-kagom\'e model (see Fig.~\ref{fig:lattice}c). It displays a smooth crossover of the flat band from the middle of the spectrum to the lowest energy. This model was already briefly introduced and its energy spectrum discussed by Asano and Hotta (see Fig.~7 in Ref. \cite{Asano2011}). Of particular interest here is the continuous deformation of the flat band without gap opening when deviating from the two limits. This deformation is accompanied by the creation of pairs of Dirac points -- i.e. linear band contact points between two bands.  

Very recently, W. Jiang et al.~\cite{WeiJiang2019} have considered the Lieb-kagom\'e model in the presence of a time-reversal breaking term (related to intrinsic spin-orbit coupling) that gaps the energy spectrum. The topology of the isolated bands was then characterized by computing the corresponding Chern numbers, which obviously depends on the way the gap is opened. In contrast, in the present article, we do not gap the energy spectrum and instead characterize the band contact points as topological defects by computing the associated charge.

On the one hand, slightly deforming the Lieb lattice, two pairs of Dirac points are generated from the three band contact at the M$=(\pi,\pi)$ point of the first Brillouin zone (BZ) (see Fig.~\ref{fig:spectrum}a and b). One pair is between the lower two bands and will evolve in the  flat band of the kagom\'e limit. The other pair (between the upper two bands) will separate and turn into the Dirac points of the kagom\'e lattice. As we will see below, near the M point, each pair has opposite winding numbers ($+-$). On the other hand, deforming the kagom\'e lattice, a single pair of Dirac points emerges from the quadratic-flat band contact at the $\Gamma$ point between the two lower bands (see Fig.~\ref{fig:spectrum}f and e). The latter has a $+2$ winding number that produces a pair of Dirac points with identical winding numbers ($++$).

The evolution of the lower two bands then raises the question of how the two distinct fusion scenarios of Dirac points can be smoothly connected, namely, a $(+-)$ pair of Dirac points emerging at M evolving into a $(++)$ pair merging at $\Gamma$. Actually, a similar phenomenon has been studied by us in a two-band model, the staggered Mielke lattice \cite{Montambaux2018}. While the integer-valued winding number characterizes the circulation of the Dirac spinor on the great circle of the Bloch sphere, the great circle itself is not fixed generally - it evolves according to the Hamiltonian parameters.
We have shown that, in order to describe properly a pair of Dirac points and their merging, the orientation of the great circle around each Dirac point has to be specified by a vector perpendicular to the plane of the great circle, that we named the \textit{winding vector}. Then during the smooth evolution  between the two merging scenarios,  the relative direction of winding vectors continuously evolves from a parallel to an antiparallel situation.

In the three-band Lieb-kagom\'e model, the mechanism of distortion and restoration of the flat band implies the generation of Dirac pairs which follows different merging scenarios. The winding vector description \textit{a priori} applies only in a two-band system. One of our main task is thus to find a consistent basis to represent the Lieb-kagom\'e Hamiltonian, local in $\k$-space, from which effective two-band descriptions for the contact points can be derived.

There is a deep connection with the work by Ahn et al. on the topological characterization of real band structures \cite{Ahn2019}. This is best seen by working in a basis in which the Bloch Hamiltonian of the Lieb-kagom\'e model is real. We detail this comparison in Appendix \ref{app:realblochham}.

The paper is organized as follow. In Sect. II, we define the model and give a brief summary of the universal Hamiltonians of Dirac points merging/emergence. In Sect. III and IV, we construct the smooth basis with which effective two-band Hamiltonians can be derived. Then we determine the evolution of the winding vectors. In Sect. V, we give a global picture of the evolving winding vector and follow with conclusions. In the Appendices~\ref{app:basis} and \ref{app:sopt}, we give details on several computations. We also generalize the Lieb-kagom\'e model in Appendix~\ref{app:glk}. In Appendix~\ref{app:realblochham}, we discuss the Lieb-kagom\'e model in an alternative representation in which the Bloch Hamiltonian is real and make contact with Ref.~\cite{Ahn2019}.

\section{Lieb-kagom\'e lattice and contact points}
\subsection{Tight-binding model}\label{model}
The Lieb-kagom\'e tight-binding model is schematically shown in Fig.~\ref{fig:lattice}c, where the full bonds indicate hopping amplitudes of fixed magnitude (which we take as energy unit) and the dashed bonds correspond to variable hopping of strength $t$. The unit cell contains three orbitals at sites A, B and C, which results in three energy bands. In the present work, we are interested in the evolution of the band structure for the parameter range $0 \leq t \leq 1$. For any $t$, this model has time-reversal symmetry, inversion symmetry (with center either on the $B$ sites or in the middle of the unit cell shown in Fig.~\ref{fig:lattice}c) and it also possesses mirror symmetries with respect to reflections through the diagonal (d) and antidiagonal (a) lines passing through B-sites of the unit cell. Two well-known limits of the above model are:

(1) the Lieb lattice $t=0$. It is a regular square lattice with B-sites forming the Bravais lattice and additional sites (A- and C-sites) located at the center in between all the nearest-neighbor B-sites (Fig.~\ref{fig:lattice}a). In addition to the above symmetries, it is also bipartite (i.e. it has a chiral sublattice symmetry) and it possesses mirror symmetries along the $x$ axis and along the $y$ axis. Since it is bipartite, its spectrum respects the energy inversion $E\rightarrow -E$ symmetry and hence the middle band is necessarily a zero-energy flat band. According to the classification of Aoki \textit{et al.}, this flat band is of topological origin \cite{Aoki1996}. For example, its energy is not affected by a perpendicular magnetic field \cite{Aoki1996}. Another example of a topological flat band is found in the dice lattice \cite{Sutherland1986}.

(2) the kagom\'e lattice $t=1$. Here it is deformed from the usual kagom\'e to have a square symmetry but it has the same lattice connectivity. The kagom\'e lattice belongs to a family of frustrated tight-binding models constructed on line graphs of bipartite parent lattices, of which the Mielke (checkerboard) lattice is another instance~\cite{Mielke,Bergman,Chalker}. These models display a flat band at the minimum (or maximum) energy which is in contact with a dispersive band. The electron localization leading to the flat band is due to a destructive interference, which is affected by a perpendicular magnetic field \cite{Aoki1996}.

\begin{figure}
\includegraphics[width=8cm]{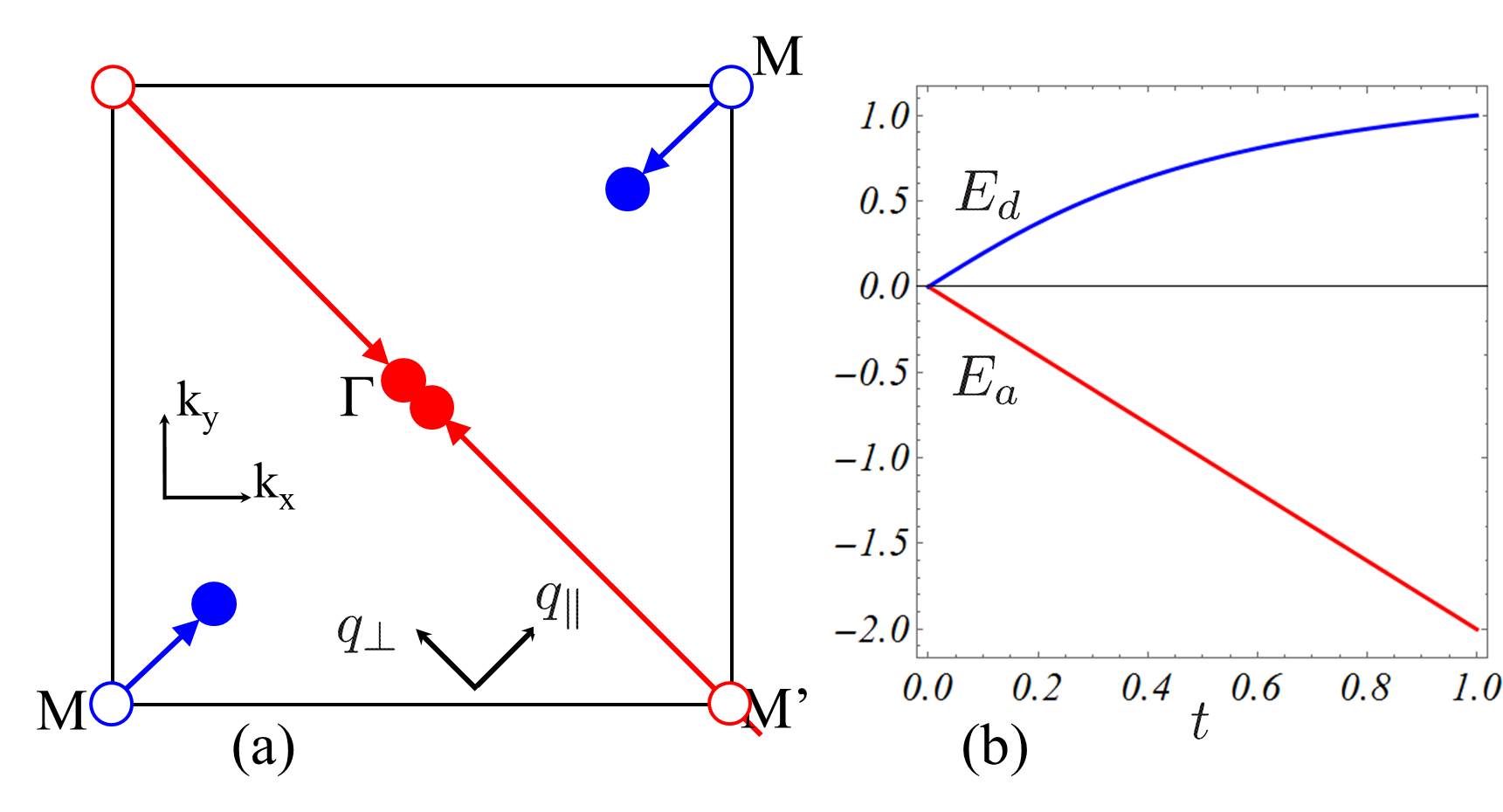}
 \caption{a) Evolution of the upper (lower) Dirac points in blue (red) in the first Brillouin zone (BZ). When $t$ increases from $0$ to $1$, the lower pair (red) moves from the $M=(\pi,\pi)$ point to the $\Gamma=(0,0)$ point, while the upper pair (blue) moves from M to $\xi \mathbf{K}=\xi (2\pi/3,2\pi/3)$ shown as filled blue disks ($\xi=\pm 1$ is a valley index). 
 b) Evolution of the energy of the upper (blue) and lower (red) Dirac points between the Lieb ($t=0$) and the kagom\'e ($t=1$) lattice.} \label{fig:BZ}
\end{figure}

We choose to write the Bloch Hamiltonian in the form $H(\k)=e^{-i \k\cdot \mathbf{R}  } H e^{i \k\cdot \mathbf{R}} $ where $\mathbf{R}$ are the positions of the Bravais lattice sites  (see Appendix \ref{app:basis} and \ref{app:basis2}). In the momentum space basis $\{ |\k,A \rangle, |\k,B\rangle,  |\k,C\rangle \}$ the Bloch Hamiltonian can be represented as 
\beq
H(\k)= \left(
  \begin{array}{ccc}
   0 & 1+e^{i k_x} & t (1+ e^{i (k_x+k_y)}) \\
    \ldots & 0 & 1+e^{i k_y} \\
   \ldots & \ldots & 0 \\
  \end{array}
\right),
\label{hamLK}
\eeq
where we have chosen the lattice spacing $a=1$ (see Fig~\ref{fig:lattice}a).

In the following, we will use the following convention for Bloch Hamiltonians. Three-band Hamiltonians in the site basis will be denoted $H$, three-band Hamiltonians in another basis $\tilde{H}$, and two-band effective Hamiltonians $\mathcal{H}$.

\subsection{Energy spectrum}
\label{enspec}
The energy spectrum for $0 \leq t \leq 1$ is shown in Fig.~\ref{fig:spectrum}, which displays three energy bands and remains gapless. The band touching points and their characteristics depend on $t$.

For the Lieb lattice ($t=0$) there is a triple band crossing at the M point in momentum space ($k_x=k_y=\pi$ and their equivalents, Fig.~\ref{fig:BZ}a). It shows a linear contact point with a flat band in the middle, which is characteristic of a pseudospin-1 \cite{Goldman2011}. In deviating from the Lieb limit ($t>0$), the flat band is distorted with the triple crossing point splitting into two pairs of Dirac points and particle-hole symmetry of the energy spectrum is lost. They are formed respectively in between the lower and middle bands, and the middle and the upper bands, in two orthogonal directions in momentum space, see Fig.~\ref{fig:zoom}.

Moving away from the Lieb limit $t>0$, the distortion of the middle band gets more pronounced and the lower Dirac pair moves in the antidiagonal direction $(k_x=-k_y)$ towards the $\Gamma$ point of the BZ ($k_x=k_y=0$) (red arrows in Fig.~\ref{fig:BZ}a). These Dirac cones are critically tilted such as to exhibit a zero-velocity line as seen on Figs.~\ref{fig:spectrum}c) and \ref{fig:zoomGamma}. In approaching the kagom\'e limit $t\rightarrow 1$, the same Dirac points \textit{merge} at the $\Gamma$ point forming a single contact point between a quadratic band and a flat band, thus restoring a flat lowest energy band. The upper Dirac pair, on the other hand, moves along the diagonal direction ($k_x=k_y$, blue arrows in Fig.~\ref{fig:BZ}a) towards a fixed position  $\xi\K=\xi(2\pi/3,2\pi/3)$, where $\xi=\pm 1$ is a valley index, while maintaining the Dirac cone structure throughout. The tilt of these Dirac cones diminishes with increasing $t$ and vanishes at $t=1$ [see Fig.~\ref{fig:spectrum}(c) and (d)].

In summary, the crossover between flat bands happens by producing pairs $(\k_D,-\k_D)$ of Dirac points. The lower pair evolves from the M and the $\Gamma$ point. Its merging/emergence at time-reversal invariant momenta (TRIM) resembles a situation we already encountered in the two-band model on a staggered Mielke lattice \cite{Montambaux2018}. There, we provided a complete scenario of a $(+-)$ going into a $(++)$ Dirac pair. In the following subsection we briefly summarize the main results of Ref.~\cite{Montambaux2018}.

\subsection{Characterization of contact points: $2\times 2$ effective Hamiltonians}
A contact point between two bands is  characterized by an effective $2\times 2$ Hamiltonian. Around a single contact point $\mathbf{k}_D$, the local two-band Bloch Hamiltonian takes on a general form
\beq
\mathcal{H} (\mathbf{q})=h_\mu(\mathbf{q})\sigma_\mu+h_\nu(\mathbf{q})\sigma_\nu \equiv\vec{h}(\mathbf{q})\cdot \vec{\sigma}
\eeq
involving only two among three Pauli matrices $\sigma_\mu=\vec{u}_\mu(\mathbf{k}_D)\cdot\vec{\sigma}$ and $\sigma_\nu=\vec{u}_\nu(\mathbf{k}_D)\cdot\vec{\sigma}$ with $|\vec{u}_{\mu,\nu}|=1$, $\vec{u}_{\mu}\cdot \vec{u}_{\nu}=0$ (e.g., $\mu,\nu=x,y$) and $\vec{\sigma}=(\sigma_x,\sigma_y,\sigma_z)$. The winding number is the number of times the effective magnetic field $\vec{h}(\mathbf{q})$ winds on a great circle of the Bloch sphere in the $(\vec{u}_\mu,\vec{u}_\nu)$ plane when $\q$ encircles the contact point once in the trigonometric direction. For example, for a Dirac point one can have $\vec{h}(\mathbf{q})=(v_x q_x,v_y q_y,0)$ and the winding number is $\text{sign}(v_x v_y)$.

Indeed, the stability of two-band contact points in 2D requires a symmetry protection (such as a chiral symmetry or inversion and time-reversal symmetries). The associated topological charge is the  $\mathbb{Z}$ winding number that classifies maps from a loop $S^1$ encircling the contact point in reciprocal space to a great circle $S^1$ of the Bloch sphere associated to the $2\times 2$ Bloch Hamiltonian $\mathcal{H}(\mathbf{q})$. In the absence of such a symmetry, band contact points are unstable defects in 2D \cite{Volovik}.

To describe a {\it pair} of Dirac points when they come close together in the vicinity of a time-reversal invariant momentum (TRIM), two merging scenarios are possible corresponding to the following two universal Hamiltonians \cite{Montambaux2009,Gail2012}
\beq {\cal H}_{+-}&=& \left(\Delta + {q_x^2 \over 2 m}\right) \,  \sigma_\mu + c q_y \, \sigma_\nu ,
\label{eq:uh+-} \\
{\cal H}_{++}&=& \left(\Delta + \frac{q_x^2  -q_y^2}{2 m} \right) \,   \sigma_\mu + { q_x q_y  \over m'}\, \sigma_\nu,
\label{eq:uh++}
\eeq
where $\sigma_\mu$, $\sigma_\nu$ are two Pauli matrices and $\Delta$, $c$, $m$ and $m'$ are parameters. Actually, because of time-reversal symmetry, there is not much choice in the Pauli matrices: in the $(+-)$ case $\sigma_\nu$ should be $\sigma_y$ and in the $(++)$ case, the $\sigma_y$ matrix should not appear in the Hamiltonian.

Making an expansion around each Dirac point, the first (resp. second) Hamiltonian describes two Dirac points with opposite (resp. equal) winding numbers located at position $(q_x,q_y)=(\xi\sqrt{-2m\Delta},0)$ if $\Delta<0$. If $\Delta>0$, Dirac points only exist in the second case and are located at  $(q_x,q_y)=(0,\xi\sqrt{2m\Delta})$. Here the winding number $\pm 1$ associated with a Dirac point is actually defined with respect to the winding plane $(\sigma_\mu,\sigma_\nu)$, which can generally be parameter dependent. This is indeed the case in  situations encountered in Ref. \cite{Montambaux2018}  and  in this paper: when varying a parameter of the Hamiltonian (here the parameter $t \in [0,1]$),  the orientation of the winding plane (defined by a great circle on the Bloch sphere) varies continuously. We shall see here that a pair of Dirac points emerges following the ${\cal H}_{+-}$ scenario with $(\sigma_\mu, \sigma_\nu)= (\sigma_z, \sigma_y)$, and merges following the ${\cal H}_{++}$ scenario with $(\sigma_\mu, \sigma_\nu)= (\sigma_z, \sigma_x)$. In order to describe the continuous evolution of the winding planes attached to each Dirac point, we recently introduced the notion of a winding vector \cite{Montambaux2018}
\beq
\vec{w}=\f{1}{2\pi}\int \vec{n}\times d\vec{n}
\eeq
with the normalized pseudospin magnetic field $\vec{n}(\mathbf{k})=\vec{h}/|\vec{h}|$. $\vec{w}$ is the unit vector normal to the oriented winding plane of $\vec n$. The winding vector is a parameter-dependent vectorial quantity, supplemental to the more familiar winding number in describing a band crossing. This notion becomes essential when describing a pair of Dirac points in close vicinity since the relative orientation of the winding plane can have physical consequences.  For a Hamiltonian of the form
\beq {\cal H} = v_x q_x \sigma_\mu + v_y q_y \sigma_\nu \ , \eeq
with $\sigma_\mu=\vec u_\mu . \vec{\sigma}$ and   $\sigma_\nu=\vec u_\nu . \vec{\sigma}$,   we simply have
\beq \vec{w}= \text{sgn}(v_x v_y) \, \vec{u}_\mu \times \vec{u}_\nu \ .
\label{windingvec2}
\eeq
In the above equation, it is implicitly assumed that $\{\vec{u}_x, \vec{u}_y, \vec{u}_z \}$ is an oriented basis.

Finally we  note that a $\sigma_0$ term with momentum dependence is important for the flatness of bands in the energy spectrum but not important for wavefunctions and the topological characterization of a band contact point.

\subsection{Reduction from three to two-band effective Hamiltonians}
In order to characterize the distortion of the flat band and the evolution of Dirac points in the three-band system, we introduce local $2\times 2$ effective Hamiltonians. In order to do this, we proceed in three steps. (1) Because the contact points move along the antidiagonal (lower bands) or the diagonal (upper bands), we will first diagonalize $H(\mathbf{k})$ along these two directions for arbitrary $t$. A great help in this process comes from the mirrors symmetries. In this way we will obtain useful eigenvector bases (see Appendix \ref{app:basis}). (2) Next, for a specific $\mathbf{k}_0$ point along these directions, we will rewrite the Bloch Hamiltonian in the eigenvector basis at $\mathbf{k}_0$ and then expand in $\mathbf{q}=\mathbf{k}-\mathbf{k}_0$ up to first or second order. In the following, $\mathbf{k}_0$ will be chosen either as a TRIM ($\Gamma$ or  M points) or one of the Dirac points $\k_D$. (3) The third step consists in projecting from three to two bands to obtain an effective two-band Hamiltonian for Dirac points. 

In the following, we study the lower (Sec. \ref{eff1}) and upper Dirac points (Sec. \ref{eff2}) in turn.

\section{Lower Dirac points}\label{eff1}
The lower Dirac points move along the antidiagonal direction $k_y=-k_x$ from M to $\Gamma$ point as $t$ is varied from 0 to 1, see Sect. \ref{enspec} and Fig.~\ref{fig:BZ}. Specifically, their positions and energy are given by
\beq
\xi \mathbf{k}_a=  \xi(2\phi_a,-2\phi_a) \text{ and } E_a=-2\cos \phi_a,
\eeq
where $\xi=\pm 1$ is a valley index and the angle $\phi_a\geq 0$ is defined by
\beq
\cos\phi_a =t.
\label{phia}
\eeq

\subsection{Basis along the antidiagonal}\label{fbas}
The Bloch Hamiltonian commutes with the mirror symmetry operator $S_a(\k)$ everywhere along the antidiagonal line ($k_y=-k_x$) in the Brillouin zone. As detailed in Appendix \ref{app:basis}, this facilitates the diagonalization of the Hamiltonian using common eigenstates with the symmetry operator, which we denote as $(|u_+\rangle,|u_-\rangle, |u_3\rangle)$. We then form the unitary matrix $L(k_x)=(|u_+\rangle\,\, |u_-\rangle\,\, |u_3\rangle)$ that allows us to write the Bloch Hamiltonian in its eigenbasis as
\beq
\tilde{H}_a(k_x,-k_x)=
\tr{diag}(E_+,E_-,E_3)
\eeq
with $E_\pm(k_x)=t\pm \Delta_a$, $E_3=E_a= - 2 t$, where $\Delta_a=\sqrt{t^2 + 8 \cos^2 \frac{k_x}{2}}$. The largest eigenvalue is $E_+$ and the second is $E_- \geq E_3$ if $t\geq \cos(k_x/2)$ and $E_3 \geq E_-$ otherwise. Along the antidiagonal line, they describe the three energy bands. At the Dirac point $k_x=2\phi_a$, the lower two bands become degenerate $E_3=E_-=-2t$ and separated from the third band $E_+=4t$. We emphasize that, while there is always a gauge freedom in the choice of the set of eigenvectors, we use here an eigenbasis which connects smoothly the two limiting cases from $t=0$ to $t=1$.

By writing the original Hamiltonian (\ref{hamLK}) in the new basis $\tilde{H}(\k)=L^\dag(k_x) H(\k) L(k_x)$, we now want to successively describe the vicinity of the M point, the $\Gamma$ point and the Dirac points (and not limited to $k_y=-k_x$), by appropriate expansions.

\subsection{Close to the Lieb limit: emergence of Dirac points at M point}
\label{sect:close-Lieb}

This is the limit $t\to 0$ when the lower Dirac points pair emerge from the triple crossing at $M=(\pi,\pi)$ point of the Lieb lattice along the antidiagonal line. Upon a momentum expansion at M point for small $\mathbf{q}$, we get $\tilde{H}_M (\q)=L(\pi)^\dag H(M+\q) L(\pi)$ in the form
\beq\label{HamM}
\tilde{H}_M (\q)=\left(\begin{array}{c|cc}
              2 t   &  i  q_\perp    & i \sqrt{2} t  q_\parallel \\ \hline
               \ldots & 0 & - i q_\parallel +\frac{q_\perp q_\parallel}{\sqrt{2}} \\
            \ldots & \ldots &     -2 t  \\
             \end{array}
           \right)
\eeq
with $q_\parallel \equiv \f{q_x+q_y}{\sqrt{2}}$ (``diagonal" direction) and $q_\perp \equiv  \f{q_y-q_x}{\sqrt{2}}$ (``antidiagonal" direction). We keep the second order in $q_\parallel$, $q_\perp$ and $t$  in the relevant blocks (indeed, we will see that the Dirac points correspond to $|q_\perp| \sim t$). The vertical/horizontal lines in the matrix merely serve as an eye guide separating the subspaces $(|u_+\rangle)$ and $(|u_-\rangle,|u_3\rangle)$ which the matrix operates. 

We then eliminate the highest band using second order perturbation theory. Following the L\"owdin method (see Appendix \ref{app:sopt}) \cite{Lowdin51} with typical energy $E_0=E_a=-2t$, we obtain an effective $2\times 2$ Hamiltonian acting in $(|u_-\rangle,|u_3\rangle)$ subspace as
\beq
\mathcal{H}_{M}(\q)&\simeq&-\bigl(  t + \f{q_\perp^2}{8t}   \bigr) \sigma_0+ q_\parallel  \sigma_y +                  \bigl(  t -   \f{q_\perp^2}{8t}   \bigr)  \sigma_z ,
\label{HMlower}
\eeq
at first order in $t$. Apart from the identity term $\sigma_0$, it has the form of the universal Hamiltonian $\mathcal{H}_{+-}$, see Eq.~(\ref{eq:uh+-}), with $\sigma_\mu=\sigma_z$ and $\sigma_\nu=\sigma_y$. It describes the merging of two Dirac points located at $(q_\parallel ,q_\perp)=(0,\xi 2 \sqrt{2}t)$ and of opposite winding numbers in the $\sigma_y-\sigma_z$ plane. The winding vectors are $\vec{w}_\xi = -\xi \vec{u}_x$. 

At second order in $t$, the effective two-band Hamiltonian (\ref{HMlower}) gets an additional contribution, which, as we show below, is responsible of a rotation of the winding vectors:
\beq
                          \frac{q_\parallel q_\perp}{2\sqrt{2}}\sigma_x .
                          \label{eq:rwv}
\eeq
Upon the substitution $q_\perp \to \xi 2\sqrt{2}t +q_\perp$, the linearized Hamiltonian near the Dirac points becomes:
\beq
\mathcal{H}_{a}(\q)\simeq\bigl(-2t -\xi \frac{q_\perp}{\sqrt{2}}  \bigr) \sigma_0+ q_\parallel  (\sigma_y +\xi t \sigma_x) -\xi  \frac{q_\perp}{\sqrt{2}}  \sigma_z.
\eeq
The Dirac cones are critically tilted because of the fine-tuning of the $q_\perp \sigma_0$ term with respect to the $q_\perp \sigma_z$ term. Also, there is some mixing between $\sigma_y$ and $\sigma_x$ indicating that the winding vectors rotate. Indeed $\vec{w}_\xi = -\xi \vec{u}_x + t \vec{u}_y$, which evolve from $-\xi\vec{u}_x$ (i.e. anti-parallel) at small $t$ to $\vec{u}_y$ (i.e. parallel) when increasing $t$.

\begin{figure}
\includegraphics[width=8.5cm]{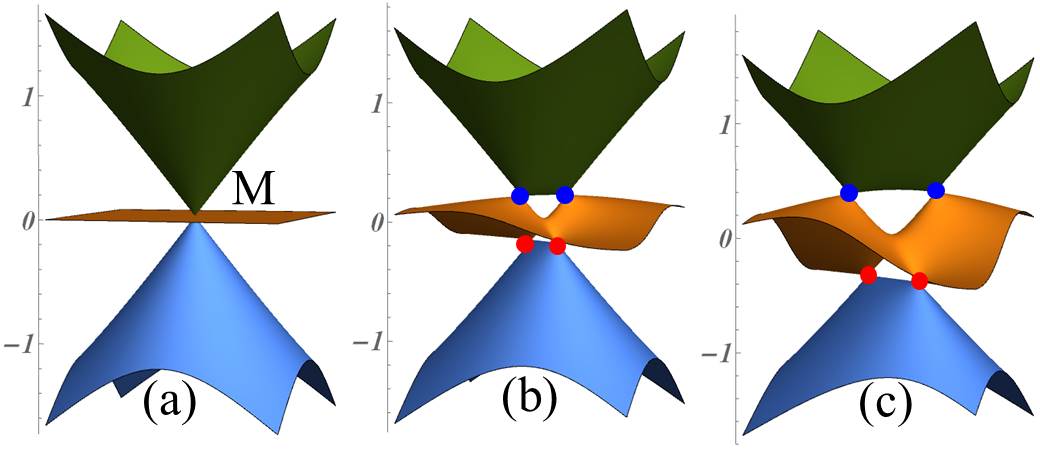}
 \caption{Close up energy spectrum at M-point showing the multiplets of Dirac points in the vicinity of the Lieb limit $t= 0 (a),0.1 (b) ,0.2 (c)$.   }\label{fig:zoom}
\end{figure}

We remark that the effective Hamiltonian near the M point has a peculiar structure. It results from an expansion \textit{a priori} valid for small $\mathbf{q}$. But near the merging when $t\rightarrow 0$, one cannot restrict to the two lowest band, since three bands become degenerate in the Lieb limit. Moreover, in this limit, the curvature of the spectrum goes to infinity, which is natural since the spectrum at the merging is indeed conical. The purpose here is to show that the Dirac points show opposite winding numbers and emerge following the $(+-)$ scenario. It should not be applied strictly at the Lieb limit $t=0$ (see the corresponding discussion in Appendix \ref{app:glk}).

\subsection{Close to the kagom\'e limit: merging of Dirac points at $\Gamma$ point}
\label{sect:close-kagome}
Close to the kagom\'e limit ($t=1$), we consider an expansion of the Hamiltonian at the $\Gamma=(0,0)$ point to give $\tilde{H}_\Gamma (\q)=L(0)^\dag H(\Gamma+\q) L(0)$ as
\beq
\tilde{H}_\Gamma (\q)= \left(
  \begin{array}{c|cc}
   4 &-i q_\perp &  i \sqrt{3} q_\parallel  \\ \hline
	\ldots &  -2-\f{2 \epsilon}{3}+\f{q_\perp^2}{3} & \displaystyle -\f{q_\parallel q_\perp}{\sqrt{3}}\\
	\ldots & \ldots &  \displaystyle  -2+2\epsilon +q_\parallel^2 
  \end{array}
\right)
\label{eq:15}
\eeq
with $t=1-\epsilon$ and $\epsilon \to 0$. Similarly to the previous section we keep up to the second order in the variables $q_\parallel$, $q_\perp$ and first order in $\epsilon$ (as we will see that the Dirac points correspond to $q_\perp^2 \sim \epsilon$) in the relevant matrix elements. Because we will treat the highest band as a perturbation to the lower sub-space, we only need to consider the off-diagonal blocks at first order and the highest band energy at zeroth order. From L\"owdin method (see Appendix \ref{app:sopt}), we arrive at the effective two-band Hamiltonian
\beq
\mathcal{H}_{\Gamma}(\q)&=&\bigl(-2+\f{2 \epsilon}{3} +\f{q_\parallel^2}{4}+\f{q_\perp^2}{12}\bigr)\sigma_0\nn\\
&&\!\!\!\!\!\!\!+\bigl(\f{q_\perp^2}{12}  -\f{q_\parallel^2}{4} -\f{4 \epsilon}{3}\bigr)\sigma_z-\f{q_\parallel q_\perp}{2\sqrt{3}} \sigma_x ,
\label{eq:14}
\eeq
at second order in $q$ and first order in $\epsilon$. Besides the identity term, it has the form of the universal Hamiltonian $\mathcal{H}_{++}$, see Eq.~(\ref{eq:uh++}) with $\sigma_\mu=\sigma_z$ and $\sigma_\nu=\sigma_x$, following the $(++)$ merging scenario. It describes the merging of two Dirac points located at $(q_\parallel,q_\perp)=(0,-\xi 4\sqrt{\epsilon})$ if $\epsilon>0$ [beware that the $+\k_a$ Dirac point is at $q_\perp=-4\sqrt{\epsilon}$] and of identical winding numbers  in the $\sigma_x-\sigma_z$ plane. Note that for $\epsilon<0$, the merging is along the $q_\parallel$ direction as the Dirac points are located at $(q_\parallel,q_\perp)=(\xi 4\sqrt{-\epsilon/3},0)$. This agrees with the model of Eq.~(\ref{eq:uh++}) showing the emergence for $t\leq 1$  as well as $t \geq 1$.

When $\epsilon=0$, this Hamiltonian describes a quadratic band crossing in the kagom\'e limit. A similar Hamiltonian was obtained by Ref. \cite{Xiao2003} in the triangular version of the kagom\'e lattice. However their Hamiltonian -- see Eq.~(2.6) in Ref.~\cite{Xiao2003} with $d_3=0$ -- breaks time-reversal symmetry which is not possible at the $\Gamma$ point. This is due to taking the Pauli matrices $\sigma_x$, $\sigma_y$ instead of $\sigma_z$, $\sigma_x$ in the Hamiltonian $\mathcal{H}_{++}$, see Eq.~(\ref{eq:uh++}).

Finding a rotation of the winding vector requires performing a computation at third order in $q\sim \sqrt{\epsilon}$ in Eq.~(\ref{eq:15}). The L\"owdin method then gives the additional terms as:
\beq
 \left(\frac{q_\perp^2-q_\parallel^2}{4\sqrt{6}} -\frac{4\epsilon}{3\sqrt{6}} \right)\,  q_\parallel \sigma_y .
\eeq

Upon the substitution $q_\perp \to -\xi 4\sqrt{\epsilon} +q_\perp$, the linearized Hamiltonian near the Dirac points becomes:
\beq
\mathcal{H}_{a}(\q)&\simeq& \bigl(-2t -\xi \f{2}{3}\sqrt{\epsilon}q_\perp\bigr)\sigma_0 -\xi \f{2}{3}\sqrt{\epsilon}q_\perp\sigma_z \nn \\
&+& \xi\, 2\sqrt{\frac{\epsilon}{3}}q_\parallel \bigl( \sigma_x +\xi \frac{2}{3}\sqrt{2\epsilon} \sigma_y  \bigl).
\eeq
The winding vector is $\vec{w}_\xi=\vec{u}_y-\xi  \frac{2}{3}\sqrt{2\epsilon} \vec{u}_x $, when $\epsilon$ is small.

\begin{figure}
\includegraphics[width=8.5cm]{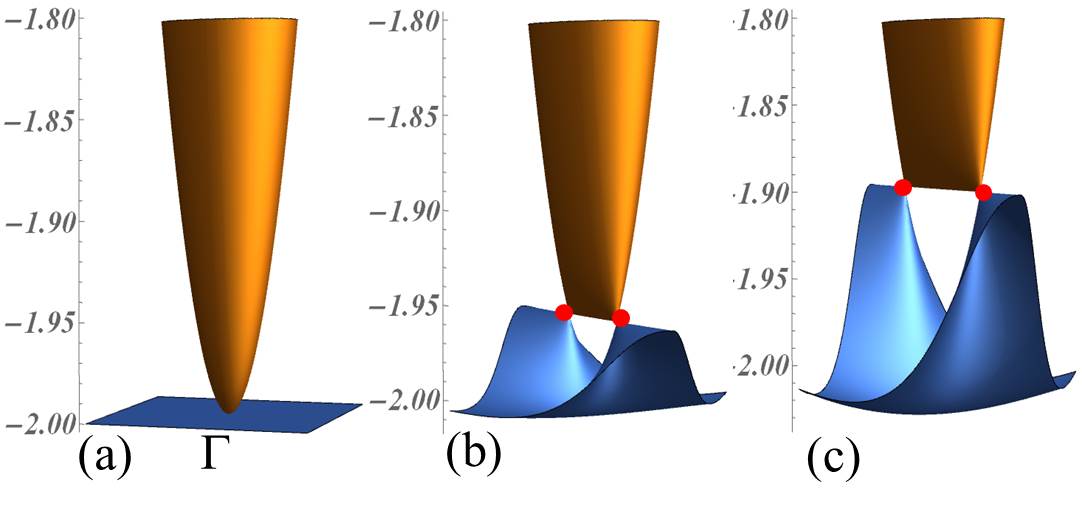}
 \caption{Close up energy spectrum at $\Gamma$-point showing the emergence of Dirac points from the kagom\'e limit $t=1 (a), 0.98 (b), 0.95 (c)$.   }\label{fig:zoomGamma}
\end{figure}

\subsection{Motion of the lower Dirac points}
We now find the paradoxical situation where a $(+-)$ pair of Dirac points near M (when $t\simeq 0$) continuously evolves into a $(++)$ pair near $\Gamma$ (when $t\simeq 1$). Naively one expects that a topological charge conservation, which holds locally in the momentum space, should also hold globally in the full BZ. Here, the global charge seems to change from $0$ (for $t=0$) to $+2$ (for $t=1$). 

To solve the apparent paradox, we now follow the evolution of the Dirac points and determine the evolution of the associated winding vector. We make an expansion following the Dirac point at $\k_a= (2\phi_a,-2\phi_a)$,
 where the angle $\phi_a$ is defined in Eq.~(\ref{phia}), to get $\tilde{H}_a(\q)=L(2\phi_a)^\dag H(\mathbf{k}_a+\mathbf{q})L(2\phi_a)$ as
\beq
\tilde{H}_a(\q)=\left( \begin{array}{c|cc} 4 t & \mathcal{O}(\q) & \mathcal{O}(\q)\\ \hline
\ldots & E_a-\f{2\sqrt{2}}{3}  \sin \phi_a q_\perp &  \sqrt{\f{2}{3}}\sin\phi_a e^{-i \phi_a}q_\parallel \\
\ldots &  \ldots  & E_a
    \end{array}    \right),
\eeq
where $E_a=-2\cos \phi_a=-2t$. For the other Dirac point at $-\k_a$, one should replace $\phi_a$ by $- \phi_a$. Because we are now interested in a Dirac Hamiltonian, which is linear in momentum, and in contrast to the two cases previously studied, we only need to keep the momentum variable up to the linear order. Corrections from the third band are necessarily of the second order and can be neglected. The effective Hamiltonian can therefore be read off from the $2\times 2$ block spanned by $(|u_-\rangle, |u_3\rangle)$ (see Appendix \ref{app:basis}) as 
\beq
\mathcal{H}_{a}(\q)\simeq \bigl(E_a- v^0_\perp q_\perp\bigr)\sigma_0 +v_\parallel q_\parallel\sigma_{\phi_a} - v_\perp q_\perp \sigma_z \ ,
\label{eq:16}
\eeq
with the (positive) velocities 
\beq
v_\parallel = \sqrt{\f{2}{3}}\sin\phi_a  \quad , \quad v_\perp =v_\perp^0= \frac{v_\parallel}{\sqrt{3}}.
\eeq
In the $t\to 0$ limit, $v_\parallel \simeq \sqrt{2/3}$ and $v_\perp =v_\perp^0\simeq \sqrt{2}/3$; whereas when $t=1-\epsilon\to 1$, $v_\parallel \simeq 2\sqrt{\epsilon/3}$ and $v_\perp =v_\perp^0\simeq 2\sqrt{\epsilon}/3$. These velocities agree with that found in the expansion near the M point (when $t\to 1$) but there is a numerical discrepancy with that found near the $\Gamma$ point (when $t\to 0$).

\begin{figure}
\includegraphics[width=8cm]{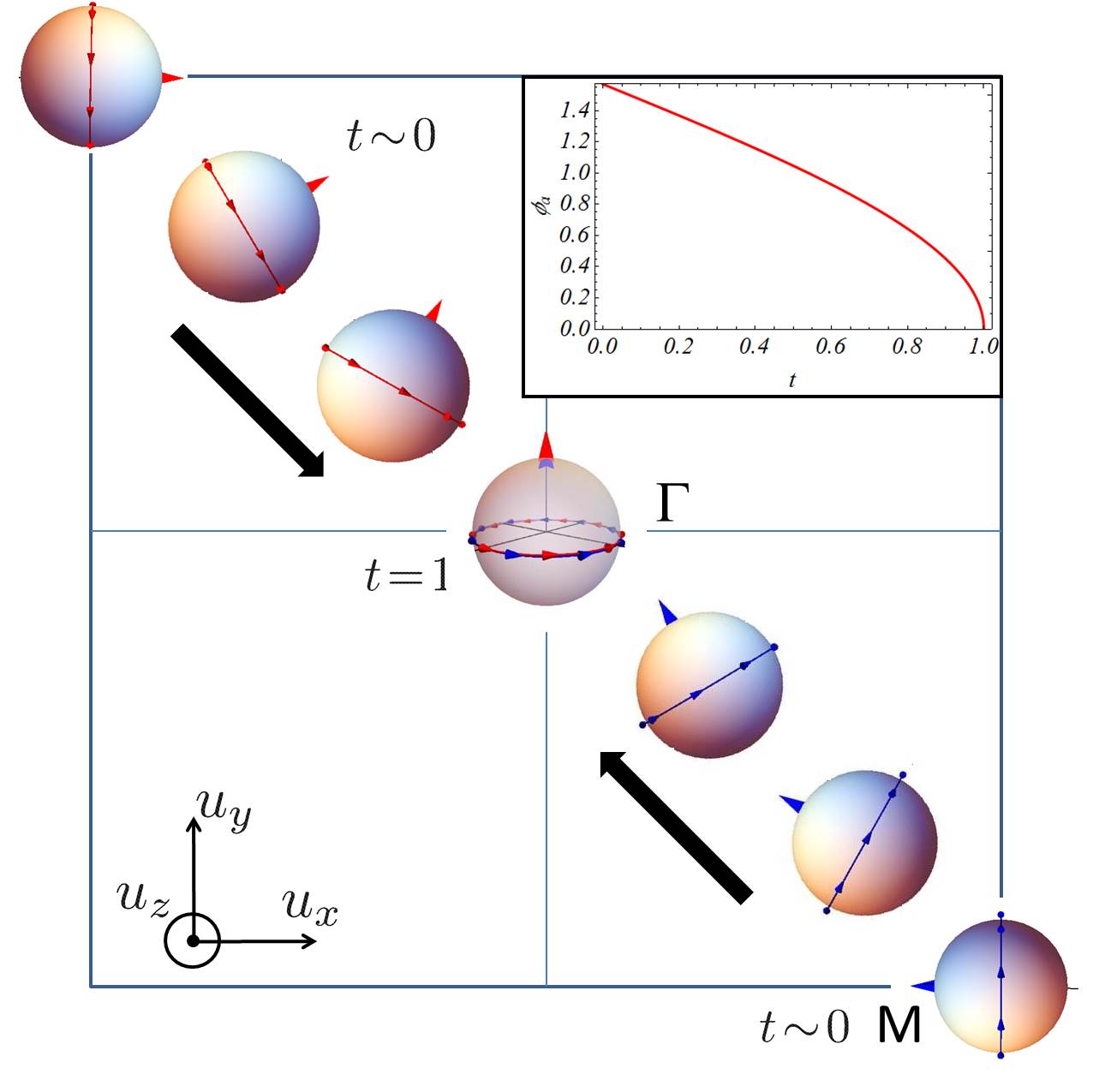}
 \caption{Winding vector evolution as a function of $t$ for the lower Dirac points. The inset shows the angle $\phi_a$ as a function of $t$. The position of Dirac points is $\xi \k_a=\xi(2\phi_a,-2\phi_a)$.}\label{fig:winding-lower}
\end{figure}

The Pauli matrix $\sigma_{\phi_a} \equiv   \cos\phi_a \sigma_x+\sin\phi_a \sigma_y$, together with $\sigma_z$, defines the pseudospin winding plane. This plane evolves in the pseudospin space from $(\sigma_y,\sigma_z)$ to $(\sigma_x,\sigma_z)$ with $\phi_a$ going from $\pi/2$ ($t=0$) to 0 ($t=1$), see Fig.~\ref{fig:winding-lower}. For the two Dirac points   at $\xi \k_a = \xi (2\phi_a,-2\phi_a)$, following Eq.~(\ref{windingvec2}),  the associated winding vectors  are:
\beq
\vec{w}_\xi&=& -\xi \sin\phi_a \, \vec{u}_x +  \cos\phi_a \, \vec{u}_y \nonumber \\
&=&-\xi \sqrt{1-t^2}\, \vec{u}_x + t \, \vec{u}_y .
\eeq
In the $t\to 0$ limit, $\vec{w}_\xi \simeq -\xi \vec{u}_x + t \, \vec{u}_y$; whereas when $t=1-\epsilon\to 1$, $\vec{w}_\xi \simeq -\xi \sqrt{2\epsilon} \vec{u}_x + \vec{u}_y$. Apart from a factor of $\frac{2}{3}$ when $t$ is close to $1$, this agrees with the results of the expansions near the M and $\Gamma$ points.
%
%

As an aside, we note that the $\sigma_0$ term modifies the dispersion of the Dirac cone quite drastically. As $v_\perp =v_\perp^0$ for all $t$, the cones are critically tilted (also known as type III Dirac-Weyl cones, see e.g. Ref.~\cite{Milicevic19} and references therein):  the velocities of excitations along the $\perp$ direction are $v_{min}=v_\perp-v_\perp^0=0$ and $v_{max}=v_\perp+v_\perp^0=2 v_\perp$. The pair of Dirac points is connected by a flat energy line (see red dots in Figs~\ref{fig:zoom} and \ref{fig:zoomGamma}). 

In summary, for the crossings in the lower two bands, Eqs.~(\ref{HMlower}) and (\ref{eq:14}) describes the emergence/merging of two Dirac cones with different winding scenarios. In the single Dirac cone Hamiltonian (\ref{eq:16}), we follow the evolving winding vector and show how the apparent winding number gets reversed as $t$ is varied.

\section{Upper Dirac points}\label{eff2}
For the upper Dirac point pair, we perform an analysis similar to the one contained in the last section. The Dirac points move along the diagonal line $k_y=k_x$, with their positions and energy (Fig. \ref{fig:BZ}) given by
\beq
\xi \k_d=\xi (2\phi_d,2\phi_d) \text{ and } E_d=2 \cos \phi_d
\eeq
and specified by the angle $\phi_d\geq 0$ such that
\beq
\cos\phi_d=\frac{\sqrt{1+8t^2}-1}{4t}.
\eeq
The upper Dirac points that emerge from the M point ($t=0$) move to a fixed position  $\xi \K=\xi (2\pi/3,2\pi/3)$ in the kagom\'e limit ($t=1$) and do not merge (in contrast to the lower Dirac points pair).

\subsection{Basis along the diagonal}
The Bloch Hamiltonian commutes with the mirror symmetry operator $S_d(\k)$ everywhere along the diagonal line ($k_y=k_x$) in the Brillouin zone. 
As detailed in Appendix \ref{app:basis}, a common basis of eigenvectors of the Hamiltonian and the symmetry operator is found, which we denote as $(|v_1\rangle,|v_+\rangle,|v_-\rangle)$. We emphasize that this basis furnishes a smooth interpolation between the two limits $t=0$ and $t=1$.
The unitary matrix is defined by $U(k_x)\equiv (|v_1\rangle\,\,|v_+\rangle\,\,|v_-\rangle)$ and written in this basis, the Hamiltonian along the diagonal line is given by
\beq
\tilde{H}_d(k_x,k_x)
=\tr{diag}(E_1,E_+,E_-)
\eeq
with $E_1=-2t\cos k_x$ and $E_\pm=t \cos k_x\pm\Delta_d$, where $\Delta_d=\sqrt{t^2\cos^2 k_x + 8 \cos^2 \frac{k_x}{2}}$. We see that the two upper bands become degenerate at energy $E_1=E_+=2\cos \phi_d =-2 t \cos(2\phi_d)$ separated from the lowest band $E_-=-4\cos \phi_d$ at the Dirac point.

\subsection{Close to the Lieb limit: emergence of Dirac points at M point}
\label{sect:close-Lieb-upper}
At $M=(\pi,\pi)$, the unitary matrix $U(\pi)=(|v_1\rangle\,\,|v_2\rangle\,\,|v_3\rangle)$ is equal to $L(\pi)=(|u_1\rangle\,\,|u_2\rangle\,\,|u_3\rangle)$ as $|v_j\rangle=|u_j\rangle$. Instead of Eq.~(\ref{HamM}), we focus on the upper two bands to obtain $\tilde{H}_M(\q)=U(\pi)^\dag H(M+\q) U(\pi)$ as 
\beq
\tilde{H}_M(\q)=\left(\begin{array}{cc|c}
              2 t   &  iq_\perp  + \frac{q_\parallel^2+q_\perp^2}{2\sqrt{2}}  & i \sqrt{2} t  q_\parallel \\
               \ldots & 0 & -i q_\parallel \\ \hline
            \ldots & \ldots &     -2 t  \\
             \end{array}
           \right),
           \label{eq:h33}
\eeq
at second order in the variables $q_\parallel$, $q_\perp$ and $t$ (as we will see that the Dirac points correspond to $q_\parallel^2 \sim t^2$) in the relevant blocks. Using L\"owdin method with typical energy $E_0=E_d\simeq 2t$, when $t\to 0$, see Appendix \ref{app:sopt}, we arrive at the effective two-band Hamiltonian (first order in $t$)
\beq
\mathcal{H}_{M}(\q)&\simeq &\big(t+\f{q_\parallel^2}{8t} \big)\sigma_0+\big(t-\f{q_\parallel^2}{8t} \big) \sigma_z  -q_\perp \sigma_y,
\label{HMupper}
\eeq
in the subspace of $(|v_1\rangle,|v_+\rangle)$. Apart from the identity $\sigma_0$ term, it takes the form of the universal Hamiltonian $\mathcal{H}_{+-}$, see Eq.~(\ref{eq:uh+-}) with $\sigma_\mu=\sigma_z$ and $\sigma_\nu=\sigma_y$, describing the emergence of two Dirac points at $(q_\parallel,q_\perp)=(-\xi 2\sqrt{2}t,0)$ [beware that the $+\k_d$ Dirac point is at $q_\parallel=-2\sqrt{2}t$] of opposite winding vectors $\vec{w}_\xi=\xi \vec u_x$ for $t\simeq 0$. The Hamiltonians (\ref{HMlower}) and (\ref{HMupper}) are similar upon exchanging the diagonal $q_\parallel$ and the antidiagonal $q_\perp$. They both describe a pair of critically tilted Dirac cones. The critical tilt is true for the lower cones but actually not for the upper ones, as seen on Fig.~\ref{fig:zoom}(c), and because of neglected higher-order terms [cf. $v_\parallel^0\neq v_\parallel$ in Eq.~(\ref{eq:velocities})].

At second order in $t$, the effective two-band Hamiltonian (\ref{HMupper}) gets an additional contribution:
\beq
\frac{q_\perp^2}{2\sqrt{2}} \sigma_x.
\label{eq:nrwv}
\eeq
This term is similar to Eq.~(\ref{eq:rwv}) except that $q_\parallel q_\perp$ is replaced by $q_\perp^2$. Whereas  Eq.~(\ref{eq:rwv}) leads to a rotation of the winding vector, it is not the case of Eq.~(\ref{eq:nrwv}). Indeed, upon the substitution $q_\parallel \to -\xi 2\sqrt{2}t +q_\parallel$, the linearized Hamiltonian near the Dirac points reads
\beq
\mathcal{H}_{d}\simeq (2t-\frac{\xi}{\sqrt{2}} q_\parallel  ) \sigma_0 - q_\perp  \sigma_y +\frac{\xi}{\sqrt{2}}  q_\parallel   \sigma_z .
\eeq
In contrast to the lower Dirac points pair, here there is no rotation of the winding vector, which is $\vec{w}_\xi = \xi \vec{u}_x$ independently of $t$. Since the Dirac cones do not approach each other again in the kagom\'e limit $t\rightarrow 1$, this concludes the analysis for double Dirac cones in the upper two bands.

\subsection{Motion of the  upper Dirac points}
To follow the evolution of a single Dirac cone, we first expand the Hamiltonian $\tilde{H}_d(\q)=U(\xi 2\phi_d)^\dag H(\k_d+\q) U(\xi 2\phi_d)$ at the position $\xi \k_{d}$ keeping linear order in $\q$. The effective Hamiltonian for the Dirac point is then obtained directly from the block Hamiltonian in the $(|v_1\rangle,|v_+\rangle)$ subspace (see Appendix \ref{app:basis}) as
\begin{equation}
\mathcal{H}_{d}(\q)
= \left(E_d - \xi v_\parallel^0   q_\parallel\right)\sigma_0
+ \xi v_\parallel   q_\parallel \, \sigma_z -  v_\perp \,q_\perp \,\sigma_y \, ,
\label{hamdeff}
\end{equation}
where $E_d=2 \cos \phi_d=\frac{\sqrt{1+8t^2}-1}{2t}$ and we defined the (positive) velocities
\beq
v_\parallel^0=\f{\sqrt{2}\sin(3 \phi_d)}{3 \cos(2\phi_d)}  \quad &,& \quad
v_\parallel=
 \f{\sin\phi_d+\sin(3\phi_d)/3}{- \sqrt{2} \cos(2 \phi_d)}  \nn \\
\text{and } v_\perp&=& \sqrt{\f{2}{3}}\sin\phi_d \ .
\label{eq:velocities}
\eeq
In the $t\to 0$ limit, $v_\parallel^0 \simeq v_\parallel \simeq \sqrt{2}/3$ (critical tilt) and $v_\perp \simeq \sqrt{2/3}$; whereas when $t\to 1$, $v_\parallel^0 \simeq 0$ (no tilt),  $v_\parallel \simeq \sqrt{3/2}$ and $v_\perp \simeq 1/\sqrt{2}$. Except when $t=0$, $v_\parallel^0\neq v_\parallel$, which means that the tilt is no critical. An effect not captured by Eq.~(\ref{HMupper}).

When $t$ varies from $0$ to $1$, the upper pair emerges from the M point and reaches two fixed points $\xi \K= \xi ({2 \pi \over 3},  {2 \pi \over 3})$ [see Fig.~(\ref{fig:BZ}a)]. Following Eq.~(\ref{windingvec2}), the winding vectors associated to the two Dirac points $\xi \k_d$ are $\vec{w}_\xi = \xi \vec u_x$. In this case, there is no rotation of the winding vectors, which remain anti-parallel.

\section{Discussion and conclusion}

We have studied in detail the evolution of Dirac points in a tight-binding model interpolating between two well-known three-bands models, the Lieb and the kagom\'e lattices. In both limits, the energy spectrum exhibits a flat band, symmetrically positioned in the middle of a conical spectrum for the Lieb lattice, at a quadratic touching with a dispersive band for the kagom\'e lattice. One of the interests of this study is to interpolate between two different flat bands: in the Lieb case, the flat band is topological (as a result of the lattice being bipartite), whereas in the kagom\'e case it results from destructive interferences. In the first case, the flat band resists the introduction of a perpendicular magnetic field, whereas it is destroyed in the second case (see the corresponding Hofstadter butterflies in Ref.~\cite{Aoki1996}).

Starting from the kagom\'e lattice, the quadratic contact point is split into two Dirac points characterized by the {\it same} winding number, following a well-characterized universal scenario. Upon further variation of the interpolating parameter, these two Dirac points merge, following another well-characterized merging scenario for a pair of Dirac points with {\it opposite} winding numbers. To solve this apparent contradiction and to characterize the topological properties of these contact points, we have reduced the full $3 \times 3$ Hamiltonian to effective $2 \times 2$ Hamiltonians using two different approaches (see below). During the evolution between the two limits, the effective Hamiltonian (or more precisely its effective pseudo-magnetic field $\vec{h}$) rotates in pseudo-spin space, leading to the notion of winding vector. The winding vectors of a pair evolve from a parallel to an anti-parallel alignment.

This work poses questions about a systematic way to obtain effective two-band Hamiltonians describing contact points in multiband models. Here, we have essentially used two different strategies: (i) an expansion of the three-band Bloch Hamiltonian at second (or third) order in the wavevector in the vicinity of a TRIM, followed by a projection (using second-order perturbation theory) in order to capture the pair of contact points located away from the TRIM; (ii) an expansion of the three-band Bloch Hamiltonian at first order in the wavevector directly in the vicinity of the contact point at $\k_D$ followed by a naive projection (without the need of perturbation theory).

Such a reduction from a multiband to an effective $2\times 2$ Hamiltonian for a pair of Dirac points is intrinsically local in $\k$ space because it is local in energy space: it can not be done in the whole Brillouin zone. It is important to realize that such a procedure carries some arbitrariness as it depends on a choice of eigenbasis. This is especially crucial for procedure (ii), which depends on an eigenbasis at the Dirac point. Indeed any linear combination of degenerate eigenvectors produces a valid alternative eigenbasis. The consequence is that the direction of the winding vector for a single Dirac point has no absolute meaning and can be changed almost at will by choosing a different basis of degenerate eigenvectors. 

However, the \textit{relative} direction between the winding vectors of the two Dirac points within a pair has an absolute meaning. This appears clearly when computing the winding vectors in the \textit{same} basis for the two Dirac points, as done in procedure (i). In other words, the effective $2\times 2$ Hamiltonian should describe both Dirac points at once. This is typically a ``universal'' Hamiltonian -- similar to $\mathcal{H}_{+-}$ or $\mathcal{H}_{++}$ -- built from a TRIM (either $\Gamma$ or M point in the present context). As we have shown, higher-order terms in momentum provide corrections to these universal Hamiltonians that can describe the rotation of the winding vectors.

In the future, it would be interesting to fully characterize a multiband problem by introducing angles that describe band couplings between more than two bands and that are valid in the complete Brillouin zone and not only locally in reciprocal space. Here, we have seen how the Lieb point splits in four Dirac points as the hopping $t$ becomes finite. It would be interesting to systematically study the different splitting scenarios of degeneracy points between three bands.

\section*{Acknowledgements}
We thank Gilles Abramovici for useful discussions on the Lieb-kagom\'e model. 
L.-K. L. is supported by the Thousand Young Talents Program of China. We thank the Institute for Advanced Study of the Tsinghua University in Beijing for hosting us while part of this work was being performed.

\appendix
\section{Basis along the antidiagonal and the diagonal}
\label{app:basis}
This appendix provides details of the derivation of the two distinct eigenbases $(|u_\pm(k_x)\rangle, |u_3(k_x)\rangle)$ and  $(|v_1(k_x)\rangle, |v_\pm(k_x)\rangle)$ that diagonalize the Bloch Hamiltonian $H(\k)$, for any value of the parameter $t$, along respectively the antidiagonal $k_x=-k_y$ and diagonal $k_x=k_y$. 
The key ingredients are the mirror symmetries $S_{a}$ and $S_d$ along the diagonal and antidiagonal respectively.

To start with, it is important to remind that the specific form of the Bloch Hamiltonian $H(\k)$ written in Eq.~(\ref{hamLK}) is a direct consequence of the implicit choice of the Bloch basis $|\k,\alpha \rangle=\sum_{\R} e^{i \k \cdot \R} |\R,\alpha \rangle$ ($\alpha=A,B,C$) that depends only on the Bravais lattice vectors $\R$ (choice known as basis I \cite{Bena2009}). The main advantage of this Bloch basis is that the Bloch Hamiltonian matrix $H(\k)$ representing the Hamiltonian is periodic $H(\k+\G)=H(\k)$ for any reciprocal lattice vectors $\G$. However an important drawback of this representation is that the matrix $H(\k)$ has lost the information of the full position of the different atoms in the unit cell. In fact, the latter information is now encoded in the matrices representing the different space symmetries of the lattice, highlighting their key role.

\subsection{Diagonalization of $H(\k)$ along the antidiagonal}\label{app:antidiag}

The Bloch matrix representation of the mirror symmetry $S_a$ along the real-space antidiagonal is given by:
\beq
S_{a}(\k)=\left(
  \begin{array}{ccc}
    0 & 0  & 1 \\
   0 & 1 & 0 \\
    1 & 0 & 0
  \end{array}
\right).
\eeq
This symmetry translates into the relation
\beq\label{sym_a}
S_a(\k)^\dag H(k_x,k_y) S_a(\k)=H(-k_y,-k_x)
\eeq
for the Bloch Hamiltonian in reciprocal space. In particular, $S_{a}(\k)$ commutes with $H(\k)$ on the antidiagonal $k_x=-k_y$. We therefore diagonalize $S_a(k_x,-k_x)$ to get the eigenvectors
\beq
|u_1\rangle&=&{1 \over \sqrt{2}}(1,0,1)^T,\nn\\
|u_2\rangle&=&(0,1,0)^T,\nn\\
|u_3\rangle&=&{1 \over \sqrt{2}}(-1,0,1)^T,
\label{v1v2v3} \eeq
with respective eigenvalues $(+1,+1,-1)$. Written in this basis, the Hamiltonian transforms into a block diagonal matrix:
\beq \tilde{H}_{a}(k_x)= \left(
  \begin{array}{cc|c}
 2 t  & 2 \sqrt{2} \cos {k_x \over 2} e^{i \frac{k_x}{2}} &  0 \\
 \ldots & 0 & 0\\ \hline
 \ldots & \ldots& -2 t
  \end{array}
\right) \qquad
\eeq
in the  $(|u_1\rangle, |u_2\rangle, |u_3\rangle)$ basis. The $2\times2$ submatrix in the subspace $|u_1\rangle,|u_2\rangle$ can be rewritten
\beq   \label{Ha_eff}
\mathcal{H}_{u_1,u_2}(k_x)= t\,  \sigma_0 + \left(
  \begin{array}{ccc}
 \Delta_a \cos \theta_a  &\Delta_a \sin \theta_a \,  e^{i \frac{k_x}{2}} \\
 \ldots & -\Delta_a \cos \theta_a \\
  \end{array}
\right)
\eeq
by introducing the parameters $\Delta_a(k_x)=\sqrt{t^2 + 8 \cos^2 \frac{k_x}{2}}$ and the angle $\theta_a(k_x)$ with $\cos \theta_a=t/\Delta_a$ and $\sin \theta_a= 2 \sqrt{2} \cos {k_x \over 2} /\Delta_a$. This has the standard form of a two-level problem with its parametric solution.

Therefore the eigenbasis and their eigenvalues which diagonalizes the Hamiltonian $\tilde{H}_{a}(k_x)$ everywhere along the anti-diagonal are
\beq\label{basisantidiag}
|u_+(k_x)\rangle &=&\cos \f{\theta_a}{2}|u_1\rangle+\sin \f{\theta_a}{2}e^{-i \frac{k_x}{2}}|u_2\rangle,\nn\\
|u_-(k_x)\rangle&=&-\sin \f{\theta_a}{2}e^{i \frac{k_x}{2}}|u_1\rangle+\cos \f{\theta_a}{2}|u_2\rangle, 
\eeq
and $|u_3\rangle$, with their corresponding eigenvalues $E_\pm(k_x)=t\pm \Delta_a$, $E_3= - 2 t$.  Finally, the chosen eigenbasis for $H(k_x,-k_x)$ reads:
\beq
|u_+\rangle &=&{1 \over \sqrt{2}} (\cos {\theta_a \over 2} , e^{-i \frac{k_x}{2}}  \sqrt{2} \sin {\theta_a \over 2} ,\cos {\theta_a \over 2} )^T ,\nn \\
|u_-\rangle &=&{1 \over \sqrt{2}} (-e^{i \frac{k_x}{2}}  \sin{\theta_a \over 2} ,\sqrt{2} \cos {\theta_a \over 2} , -e^{i \frac{k_x}{2}}  \sin {\theta_a \over 2} )^T \nn \\
|u_3\rangle&=&{1 \over \sqrt{2}}(-1,0,1)^T .
\eeq

\subsection{Diagonalization of $H(\k)$ along the diagonal}
\label{app:diag}
We now proceed very similarly along the diagonal $k_x=k_y$. The Bloch matrix representation of the mirror symmetry $S_d$ along the real-space diagonal is given by:
\beq
S_{d}(\k)=\left(
  \begin{array}{ccc}
    0 & 0  & e^{i k_x} \\
   0 & 1 & 0 \\
    e^{- i k_y} & 0 & 0
  \end{array}
\right).
\eeq
This symmetry translates into the relation
\beq\label{sym_a}
S_d(\k)^\dag H(k_x,k_y) S_d(\k)=H(k_y,k_x)
\eeq
for the Bloch Hamiltonian in reciprocal space. In particular, $S_{d}(\k)$ commutes with $H(\k)$ on the diagonal $k_x=k_y$. We therefore diagonalize $S_d(k_x,k_x)$ to get the eigenvectors
\beq
|v_1\rangle&=&e^{i s\frac{\pi}{2}} {1\over \sqrt{2}} (-e^{i \frac{k_x}{2}},0,e^{-i \frac{k_x}{2}})^T  , \\
|v_2\rangle&=& (0,1,0)^T,\\
|v_3\rangle&=&e^{i s\frac{\pi}{2}}  {1 \over \sqrt{2}}  (e^{i \frac{k_x}{2}},0,e^{-i \frac{k_x}{2}})^T  ,
\eeq
with respective eigenvalues $(-1,+1,+1)$, where $s$ is defined as $s=\text{sign } k_x$ if $0<|k_x|\leq \pi$ and $s=0$ if $k_x=0$. The reason for this phase choice is that we want the kets $|v_j\rangle$ to be real at $\Gamma$ and M, to be periodic with the BZ and that $|v_j(-k_x)\rangle = (|v_j(k_x)\rangle)^*$. Written in this basis, the Hamiltonian thus transforms into a block diagonal matrix:
\beq
\tilde{H}_{d}(k_x)=\left(
  \begin{array}{c|cc}
 -2 t \cos k_x  &0 &  0 \\ \hline
 \ldots& 0 &  e^{i s\frac{\pi}{2}}  2 \sqrt{2} \cos {k_x \over 2} \\
 \ldots & \ldots&  2 t \cos k_x
  \end{array}
\right)
\eeq
in the $(|v_1\rangle,|v_2\rangle,|v_3\rangle)$ basis. Consider the $2\times2$ effective submatrix $\mathcal{H}_{v_2,v_3}(k_x)$ in the subspace $(|v_2\rangle,|v_3\rangle)$; we obtain an effective Hamiltonian of the same structure as in Eq.~(\ref{Ha_eff}) with the substitution $\Delta_a(k_x) \rightarrow \Delta_d(k_x)=\sqrt{t^2 \cos^2 k_x + 8 \cos^2 \frac{k_x}{2}}$
and the angle $\theta_a(k_x) \rightarrow \theta_d(k_x)$ with $\cos \theta_d=-t \cos k_x /\Delta_d \ge 0$ and $\sin \theta_d= 2 \sqrt{2} \cos {k_x \over 2} /\Delta_d$, and the coefficient for the identity term $t \rightarrow t \cos k_x$. Thus, a basis that diagonalizes the Hamiltonian $\tilde{H}_{d}(k_x)$ everywhere along the diagonal is
\beq
|v_+(k_x)\rangle &=&
   \cos {\theta_d \over 2} | v_2 \rangle +
    \sin {\theta_d \over 2}  e^{-i s\frac{\pi}{2}}  | v_3 \rangle,\nn\\
		|v_-(k_x)\rangle &=&
    - \sin {\theta_d \over 2}  | v_2 \rangle +
    \cos{\theta_d \over 2} e^{-i s\frac{\pi}{2}}  | v_3 \rangle,
\eeq
and $|v_1\rangle$, with the eigenvalues $E_{\pm} =t \cos k_x \pm \Delta_d$ and $E_1=-2t \cos k_x$. Finally, the chosen eigenbasis for $H(k_x,k_x)$ reads:
\beq
|v_1\rangle&=&{ e^{i s\frac{\pi}{2}} \over \sqrt{2}} (-e^{i \frac{k_x}{2}},0,e^{-i \frac{k_x}{2}})^T  , \\
|v_+\rangle &=&{1 \over \sqrt{2}} (e^{i \frac{k_x}{2}}  \sin {\theta_d \over 2} ,\sqrt{2} \cos {\theta_d \over 2} ,e^{-i \frac{k_x}{2}}  \sin {\theta_d \over 2} )^T ,\nn \\
|v_-\rangle &=&{1 \over \sqrt{2}} (e^{i \frac{k_x}{2}}  \cos {\theta_d \over 2} ,-\sqrt{2} \sin {\theta_d \over 2} ,e^{-i \frac{k_x}{2}}  \cos {\theta_d \over 2} )^T .\nn
\eeq

\section{L\"owdin partitioning method for the effective $2\times 2$ Hamiltonian}
\label{app:sopt}
The effective Hamiltonian in a sub-space can be obtained using the method of L\"owdin \cite{Lowdin51}, also called partitioning. Let the Hilbert space be described as the direct sum of two sub-spaces labelled $\alpha$ and $\beta$. The Hamiltonian then takes the following block form:
\beq
H=\left(
  \begin{array}{c|c}
   H_{\alpha \alpha} & H_{\alpha \beta} \\ \hline
  H_{\beta \alpha} & H_{\beta \beta}  \\
  \end{array}
\right)
\eeq
It is then easy to show that the eigen-equation $H|\psi\rangle=E|\psi\rangle$ reads
\beq
\left[H_{\alpha \alpha} + H_{\alpha \beta } \left(E-H_{\beta \beta}\right)^{-1} H_{\beta \alpha}\right] |\psi_\alpha\rangle = E |\psi_\alpha\rangle
\label{eq:b6}
\eeq
in the $\alpha$ sub-space, where $|\psi\rangle=(|\psi_\alpha\rangle,|\psi_\beta\rangle )^T $. The operator in the left-hand side of the above equation acts as an effective Hamiltonian in the $\alpha$ sub-space, except that it depends on the energy $E$, i.e. (\ref{eq:b6}) is a self-consistent equation in the spirit of the well-known Brillouin-Wigner perturbation theory. Replacing $E$ in the left-hand side by a typical relevant energy $E=E_0$ gives an effective Hamiltonian
\beq
\mathcal{H}_\alpha=H_{\alpha \alpha} + H_{\alpha \beta } \left(E_0-H_{\beta \beta}\right)^{-1} H_{\beta \alpha},
\eeq
so that the approximate eigen-equation replacing (\ref{eq:b6}) reads $\mathcal{H}_\alpha  |\psi_\alpha\rangle \simeq E |\psi_\alpha\rangle$. 
Compared to a naive projection $\mathcal{H}_\alpha\simeq H_{\alpha \alpha}$, this effective Hamiltonian includes corrections $H_{\alpha \beta } \left(E_0-H_{\beta \beta}\right)^{-1} H_{\beta \alpha}$ akin to second-order perturbation theory.

Specifically, for the $3\times 3$ Hamiltonian 
\beq
H=\left(
  \begin{array}{cc|c}
   E_1 & a & c \\
    a^* & E_2 & b \\ \hline
   c^*& b^* & E_3 \\
  \end{array}
\right), \eeq
we get the effective $2\times 2$ Hamiltonian after eliminating the third band which reads
\beq
\mathcal{H}_\alpha=\left(
  \begin{array}{cc}
   \tilde{E}_1 & \tilde{a}  \\
    \tilde{a}^* & \tilde{E}_2
  \end{array}
\right)
\label{eq:halpha}
\eeq
with 
\beq
&&\tilde{E}_1=E_1+\f{|c|^2}{E_0-E_3}, \tr{\ \ }\tilde{E}_2=E_2+\f{|b|^2}{E_0-E_3} \nonumber \\
&&\tilde{a}=a+\f{c b^*}{E_0-E_3}.
\eeq
With this method there is some arbitrariness in the choice of $E_0$. In the present paper, we are interested in describing Dirac points as seen from the nearest TRIM (e.g., $\Gamma$ or M point in the BZ): this means that $E_0$ will be chosen to be the energy of the Dirac points -- called $E_a$ and $E_d$ in the main text -- in order to properly describe the vicinity of the Dirac points and not the vicinity of the TRIM.

\section{Generalized Lieb-kagom\'e model}
\label{app:glk}
\begin{figure}[h!]
\begin{center}
\includegraphics[width=3cm]{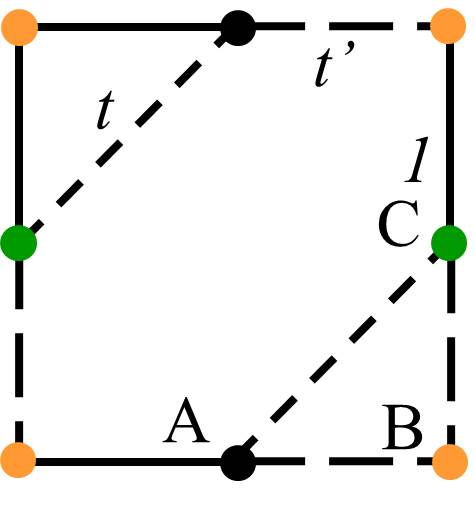}
\includegraphics[width=8.5cm]{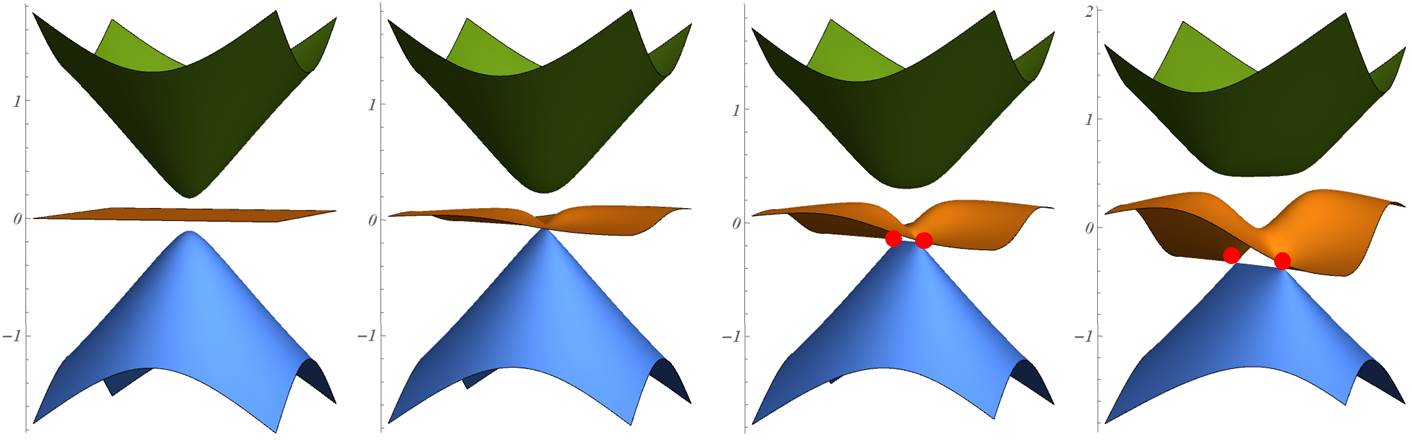}
\end{center}
 \caption{Top : dimerization of the Lieb-kagom\'e lattice. When $t' \neq 1$, a gap opens between the upper bands. Bottom : close up energy spectrum at M-point, when the upper band has been gapped. The evolution of the lower pair of Dirac points is properly described by an effective $2 \times 2$ Hamiltonian ${\cal H}_{+-}$ up to their merging. Here we have chosen  $t'=0.9$,  $t= 0,0.05,0.1,0.2$.  The merging transition is reached when $t=(1-t')/2$, here when $t=0.05$. Compare with the gapless case on Fig. \ref{fig:zoom}.} \label{fig:zoom-gapped}
\end{figure}

The effective $2 \times 2$  Hamiltonian (\ref{HMlower}) describes the evolution of the two lower Dirac points along the antidiagonal, close to the Lieb limit (red points on Fig.~\ref{fig:zoom}). Similarly the Hamiltonian (\ref{HMupper}) describes the evolution of the two upper Dirac points along the diagonal (blue points on Fig~ \ref{fig:zoom}). However, when $t \rightarrow 0$, a two-band description is not appropriate, since the {\it four} points merge together. The Hamiltonians (\ref{HMlower}, \ref{HMupper}) provide a correct description of the merging of each pair close to the M point, {\it but not too close}. They do not correctly describe the $t=0$ limit. This is obvious when considering the divergence of the $q_\perp^2/8t$ and $q_\parallel^2/8t$ terms respectively in Eqs. (\ref{HMlower}) and (\ref{HMupper}). In order to show that the merging of the two lower Dirac points is actually properly described by the ${\cal H}_{+-}$ Hamiltonian of Eq.~(\ref{HMlower}), we reconsider the same problem by gapping the upper band from the two lower bands. This is done by adding a dimerization along vertical and horizontal bonds (which breaks the mirror symmetry with respect to the diagonal and the inversion center), as shown in Fig. \ref{fig:zoom-gapped}(Top). We introduce a parameter $t' \neq 1$ on the $A-B$ bond while the coupling is still $1$ on the $B-A$ bond. The Bloch Hamiltonian (basis I) is
\beq
H(\k)= \left(
  \begin{array}{ccc}
   0 & t'+e^{i k_x} & t (1+ e^{i (k_x+k_y)}) \\
    \ldots & 0 & t'+e^{i k_y} \\
   \ldots & \ldots & 0 \\
  \end{array}
\right),
\eeq
We call it the generalized Lieb-kagom\'e model with three hopping amplitudes. This generalization is inspired by the dimerization of the Lieb lattice used in \cite{Poli2017} to gap the three bands of the Lieb model. It is also close to, but different from, the anisotropic kagom\'e model discussed in \cite{Asano2011}.

Using the same procedure as in section \ref{sect:close-Lieb}, we construct the   $2 \times 2$  effective Hamiltonian:
\beq
\mathcal{H}_{M}(\q) 
&=&-\bigl( { t + \delta \over 2} + \f{{t'}^2 q_\perp^2}{2 \delta+6t}   \bigr) \sigma_0  + v_\parallel q_\parallel  \sigma_y  \nonumber \\
&+&  \bigl(  {3 t - \delta \over 2} -  \f{{t'}^2 q_\perp^2}{2\delta+6t}   \bigr)  \sigma_z , 
\eeq
with
\beq
\delta= \sqrt{t^2 + (1-t')^2}  \,\, ,   \qquad    \tan \theta= \sqrt{2} {1 - t' \over t}   \ , \eeq
and the velocity
\beq v_\parallel = t' \cos {\theta \over 2}+ \sqrt{2} t \sin {\theta \over 2}  \ .  \eeq
The parameters have the following limits:
\begin{eqnarray}t'=1 &\rightarrow& \delta=t \, , \ \theta=0 \ , \ v_\parallel=1    \\
t=0 &\rightarrow& \delta=1 - t'   \ , \ \theta={\pi \over 2}\ , \ v_\parallel= \frac{t'}{\sqrt{2}}     \end{eqnarray}
The limit $t \rightarrow 0$ is now well behaved  when $t' \neq 1$, since the parameter $\delta$ stays finite up to the merging transition which is reached when $\delta= 3 t$, that is $t=(1-t')/2$. The full evolution is shown on Fig. \ref{fig:zoom-gapped}

\section{Real Bloch Hamiltonian and orientability}
\label{app:realblochham}
In this appendix, we study Bloch Hamiltonians that are real and discuss the notion of orientability as defined in section VI.B of Ref.~\cite{Ahn2019}. Real Bloch Hamiltonians are possible when the system has both space inversion and time-reversal symmetries. For pedagogical purposes, we start with the two-band staggered Mielke model \cite{Montambaux2018} before turning to the three-band Lieb-kagom\'e model.

We first need to introduce two different Bloch Hamiltonians that can be defined for any tight-binding model possessing more than a single site per unit cell. On the one hand, the Bloch Hamiltonian in the so-called basis I \cite{Bena2009} is defined as
\beq
H_{I}(\mathbf{k})=e^{-i \mathbf{k}.\mathbf{R}}He^{i \mathbf{k}.\mathbf{R}},
\eeq
where $\mathbf{R}$ is the Bravais lattice position operator. It is periodic with any Bravais lattice vector $\G$: 
\beq
H_I (\k+\G)=H_I (\k). 
\eeq
This basis I is also the convention we choose in the core of the article when studying the Lieb-kagom\'e model. On the other hand, the Bloch Hamiltonian in the so-called basis II \cite{Bena2009} is defined as
\beq
H_{II}(\mathbf{k})=e^{-i \mathbf{k}.\mathbf{r}}He^{i \mathbf{k}.\mathbf{r}}=e^{-i \mathbf{k}.\boldsymbol{\delta}}H_I(\k)e^{i \mathbf{k}.\boldsymbol{\delta}},
\eeq
where $\mathbf{r}=\mathbf{R}+\boldsymbol{\delta}$ is the complete position operator involving the Bravais lattice position $\mathbf{R}$ and the intra-cell position operator $\boldsymbol{\delta}$. For certain Bravais lattice vectors $\G$, it is not periodic but obeys: 
\beq
H_{II}(\k+\G)=e^{-i \G\cdot \boldsymbol{\delta}} H_{II}(\k) e^{i \G\cdot \boldsymbol{\delta}}.
\label{nphk}
\eeq

\subsection{Mielke model with  a real Bloch Hamiltonian}
\label{app:real2bands}
Here we discuss the case of a two-band real Bloch Hamiltonian, having in mind the staggered Mielke model \cite{Montambaux2018}. This model has inversion and time-reversal symmetries and the inversion center can be chosen on-site. It is defined on the checkerboard lattice. The Bloch Hamiltonian is real and non-periodic in basis II, whereas it is complex and periodic in basis I (it involves the three Pauli matrices). In basis II, under translation by a Bravais lattice vector $\G$, it behaves as  Eq.~(\ref{nphk}). When $\G$ is a basis vector of the reciprocal lattice then $e^{i \G\cdot \boldsymbol{\delta}}=\sigma_z $. The fact that $\det \sigma_z = -1$ means that the Bloch Hamiltonian is non-orientable (see section VI.B in Ref.~\cite{Ahn2019}). In other words, it is not possible to find a representation in which the Bloch Hamiltonian would be both real and periodic with the first BZ.

\begin{figure}
\begin{center}
\includegraphics[width=6cm]{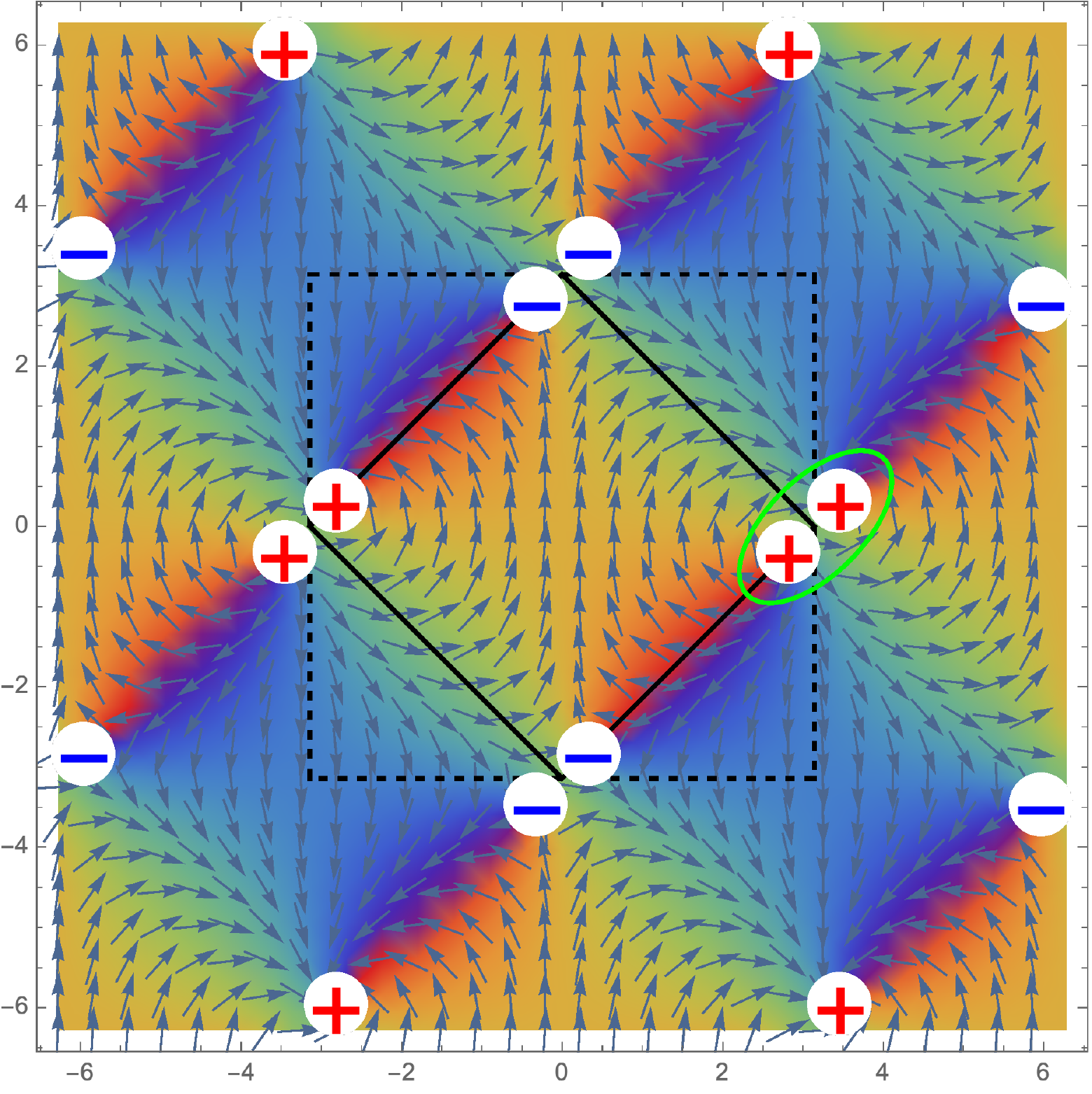}
\includegraphics[width=6cm]{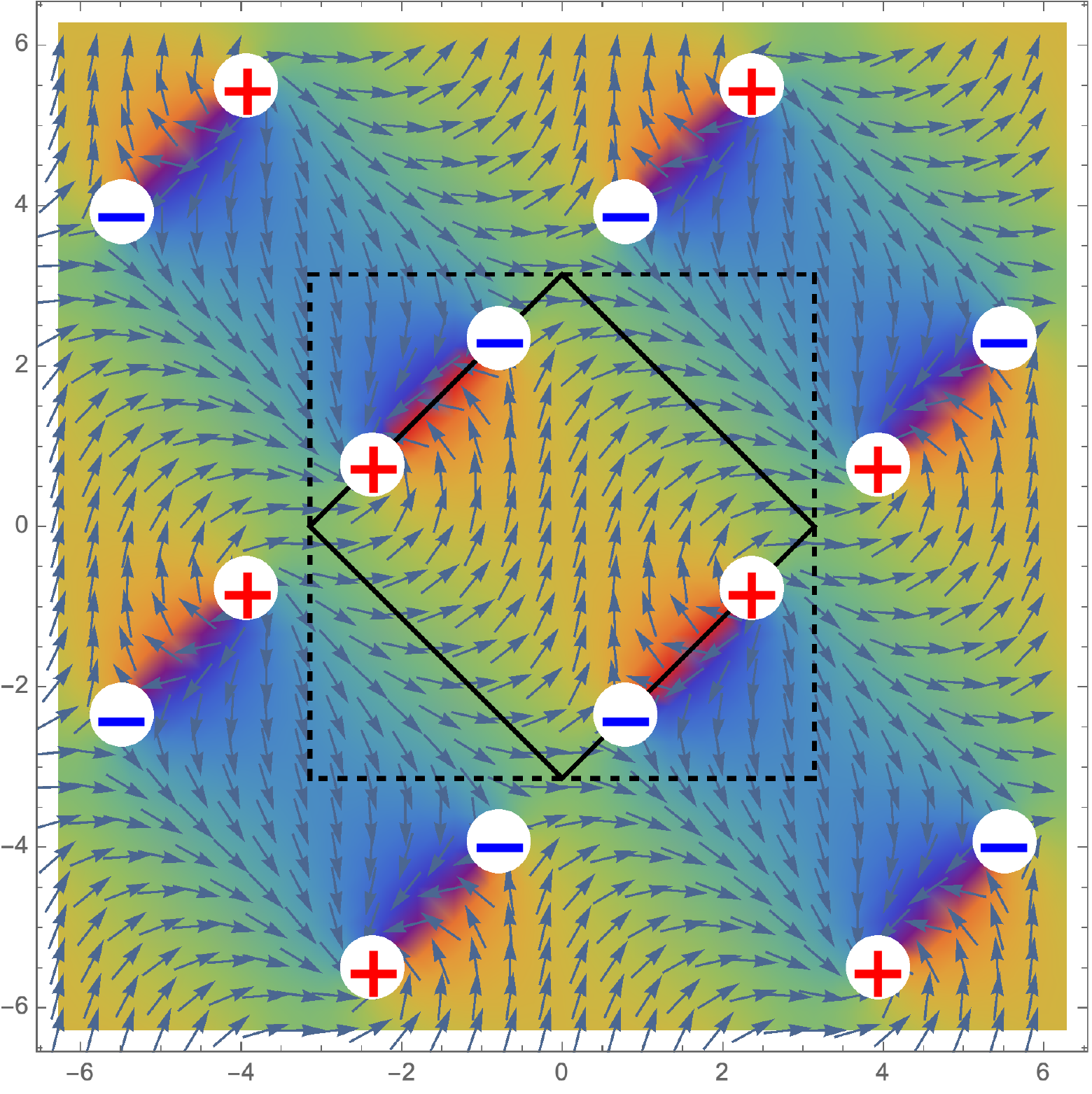}
\includegraphics[width=6cm]{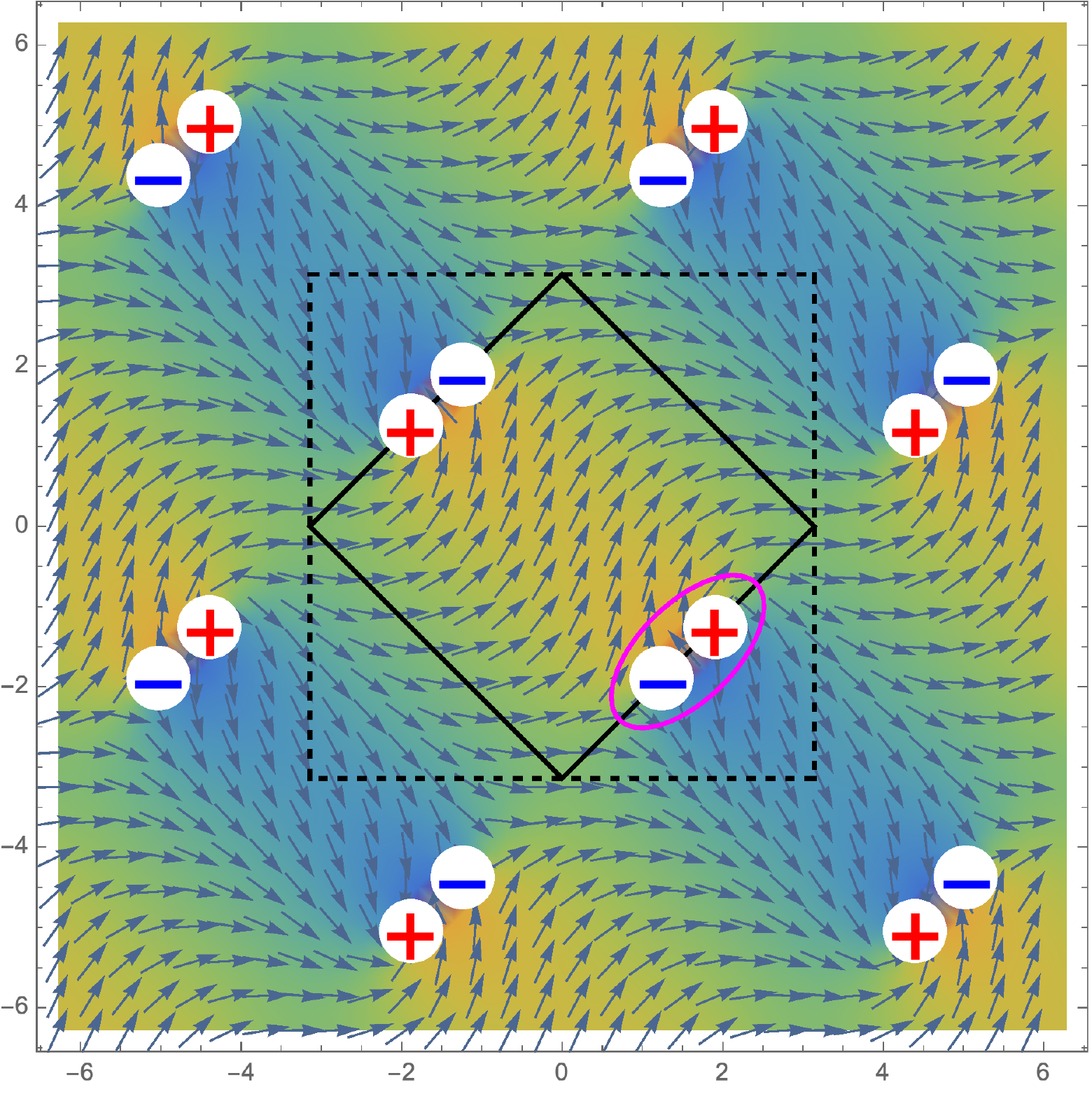}
\end{center}
\caption{Pattern of winding numbers (charges shown as red ``+" and blue ``-") for the staggered Mielke model in reciprocal space. Top: $\delta=0.1$, middle: $\delta=0.5$ and bottom: $\delta=0.9$. The first BZ is shown in black. The unit cell for the charge pattern is shown as a dashed line (it is twice larger than the first BZ). The arrows and the corresponding color map indicate the angle $\theta(\k)$ needed to define the Bloch Hamiltonian in basis II (see text). The unit cell for the Bloch Hamiltonian is the same as for the charge pattern. The green (resp. magenta) ellipse indicates the $++$ emergence (resp. $+-$ merging).}
\label{fig:chargepatternmielke}
\end{figure}

The real Bloch Hamiltonian can be written as
\beq
H_{II}(\k)&=&h_x(\k)\sigma_x+h_z(\k)\sigma_z \nonumber \\
&=&\sqrt{h_x^2+h_z^2}(\sin \theta \sigma_x +\cos\theta \sigma_z),
\eeq
which defines an angle $\theta(\k)$. It produces a checkerboard pattern in reciprocal space with a unit cell twice larger than the first BZ (see the yellow and blue zones in Fig.~\ref{fig:chargepatternmielke}). This is also the case of the pattern of topological charges (winding numbers) as show by plusses and minuses in the same figure.

Orientability of the Bloch Hamiltonian is related to being able to find a pattern of topological charges which has the periodicity of the first BZ. There is an obstruction to that in the Mielke model (see Fig.~\ref{fig:chargepatternmielke}). Fundamentally, it is this obstruction that allows the evolution of topological charges (i.e. the ``++ $ \to $ +-'' phenomenon).

\subsection{Lieb-kagom\'e model with a real Bloch Hamiltonian}
\label{app:basis2}
As an alternative to the core of the paper, we present here the evolution of the topological charges of band contact points by writing the Bloch Hamiltonian of the Lieb-kagom\'e model in basis II instead of basis I \cite{Bena2009}, as was done in Eq.~(\ref{hamLK}). In basis II, the Bloch Hamiltonian reads:
\beq
H_{II}(\k)= \left(
  \begin{array}{ccc}
   0 & 2\cos\frac{k_x}{2} & 2 t \cos\frac{k_x+k_y}{2} \\
    \ldots & 0 & 2\cos\frac{k_y}{2} \\
   \ldots & \ldots & 0 \\
  \end{array}
\right).
\eeq
This Bloch Hamiltonian turns out to be real as a consequence of time-reversal symmetry $H_{II}(-\mathbf{k})^*=H_{II}(\mathbf{k})$ and inversion symmetry $H_{II}(-\mathbf{k})=H_{II}(\mathbf{k}) $ with an on-site inversion center. In basis I, space inversion acts in a different way on the Bloch Hamiltonian and does not force it to be real.

\begin{figure}
\begin{center}
\includegraphics[width=5cm]{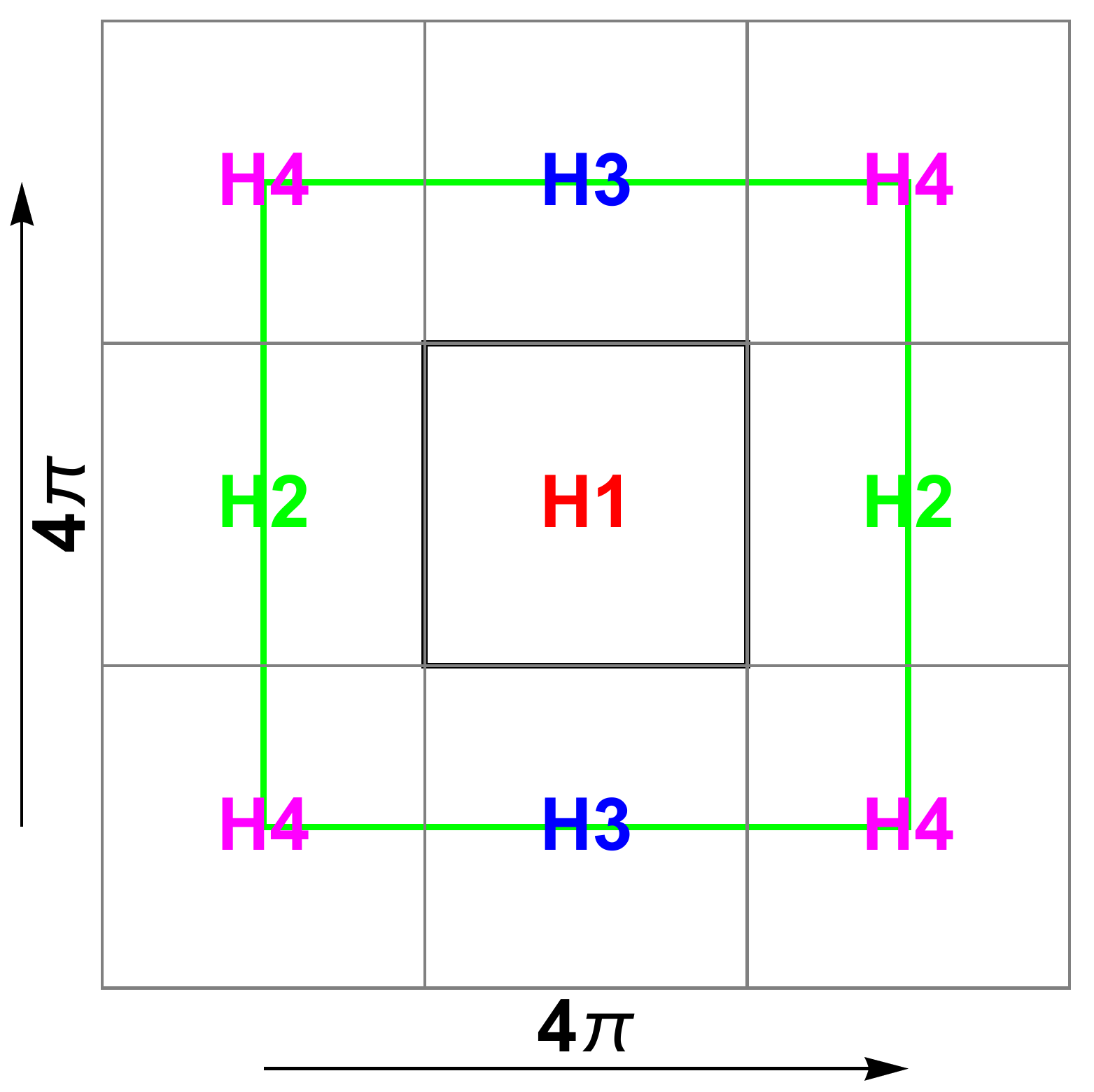}
\end{center}
\caption{Pattern of Bloch Hamiltonians for the Lieb-kagom\'e model (in basis II). The first BZ is shown as a thick black square. H1 means $H_{II}(\k)$, H2 = $H_{II}(\k+2\pi \mathbf{u}_x)$, H3 = $H_{II}(\k+2\pi \mathbf{u}_y)$ and H4 = $H_{II}(\k+2\pi \mathbf{u}_x+2\pi \mathbf{u}_y)$. The unit cell for the Bloch Hamiltonian (shown as a green square) is four times larger than the first BZ.}
\label{fig:hamiltonianpattern}
\end{figure}

In contrast to $H_I (\mathbf{k})$ [see Eq.~(\ref{hamLK})] that satisfies $H_I (k_x+2\pi,k_y)=H_I (k_x,k_y)=H_I (k_x,k_y+2\pi)$, the Bloch Hamiltonian $H_{II}(\mathbf{k})$ does not have the periodicity of the reciprocal lattice but a doubled-periodicity in both $k_x$ and $k_y$ such that $H_{II}(k_x+4\pi,k_y)=H_{II}(k_x,k_y)=H_{II}(k_x,k_y+4\pi)$ and $H_{II}(k_x+2\pi,k_y+2\pi)=-H_{II}(k_x,k_y)$. See Figure~\ref{fig:hamiltonianpattern}.

Because $H_{II}(\mathbf{k})$ is real, it means that any local two-band Hamiltonian $\mathcal{H}_{II}(\mathbf{q})$ obtained after projection in the vicinity of a band contact point is also real. Therefore $\mathcal{H}_{II}(\mathbf{q})$ can be decomposed onto $\sigma_x$ and $\sigma_z$ only, so that the corresponding contact point has a winding vector $\vec{w}=w\vec{u}_y$ and the notion of winding number $w$ is sufficient. This already means that, in basis II, there is no such thing as a continuous evolution of the winding vector.

\begin{figure}
\begin{center}
\includegraphics[width=4.2cm]{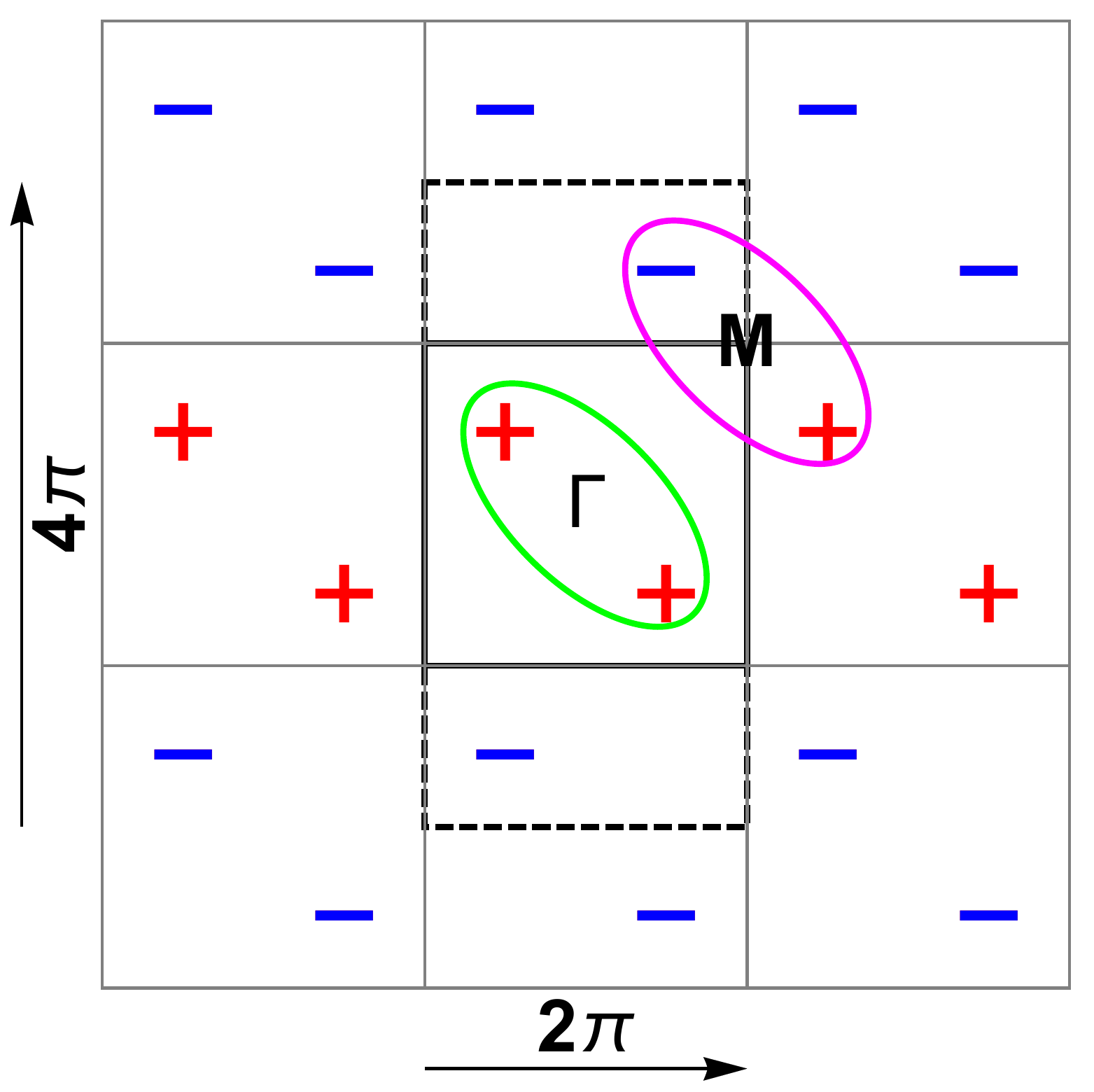}
\includegraphics[width=4.2cm]{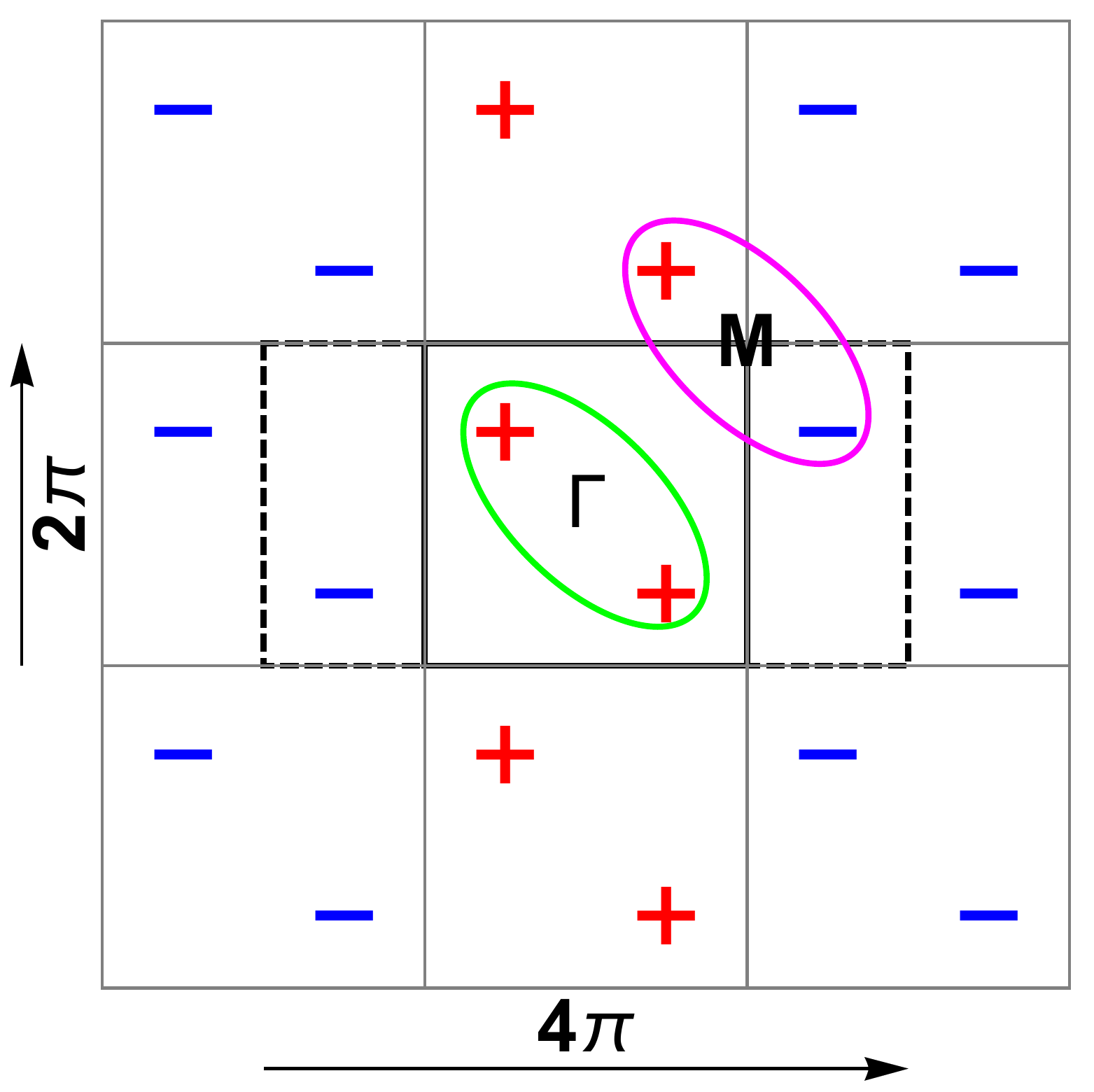}
\end{center}
\caption{Pattern of winding numbers (charges shown as red ``+" and blue ``-") for the band contact points between the lower bands in reciprocal space beyond the first BZ (shown as a thick square) as found using the Bloch Hamiltonian in basis II. The green (resp. magenta) ellipse indicates the $++$ (resp. $+-$) pattern close to the $\Gamma$ (resp. M) point. Left: one possible choice, with double periodicity along $k_y$. Right: another possibility, with double periodicity along $k_x$. The unit cell for the winding numbers (shown as a dashed rectangle) is twice larger than the first BZ.}
\label{fig:windingpattern}
\end{figure}

We only consider the contact points between the lower two bands. We find that, close to the kagom\'e limit and by expanding in the vicinity of the $\Gamma$ point (similar to section \ref{sect:close-kagome}), the pair of Dirac point has identical winding numbers (either $++$ or $--$ depending on the basis choice at $\Gamma$). In the Lieb limit and by expanding in the vicinity of the M point (similar to section \ref{sect:close-Lieb}), we find that the pair of Dirac points has opposite winding numbers (either $+-$ of $-+$ depending on the basis chosen at M). How is this possible? Actually, the Bloch Hamiltonian does not have the periodicity of the reciprocal lattice but a doubled periodicity. It means that the topological charges of the contact points need not have the periodicity of the reciprocal lattice. In other words, the charges attributed to the band contact points within the first BZ are not necessarily the same in ``another BZ''. Possible patterns respecting the above rules (same charges near $\Gamma$ opposite charges near M) are shown in red in Fig.~\ref{fig:windingpattern}. The unit cell for the winding numbers is two times larger than the first BZ.

It is not enough to have a Bloch Hamiltonian with enlarged periodicity compared to the BZ in order to have a pattern of topological charges for contact points with an enlarged periodicity. Actually, the periodicity of the Bloch Hamiltonian and of the pattern of topological charges need not be the same (the unit cell in reciprocal space is four times larger than the BZ for the Bloch Hamiltonian, see Fig.~\ref{fig:hamiltonianpattern}, and twice larger for the charges, see Fig.~\ref{fig:windingpattern}). There is an extra thing that is needed. Ahn et al. have identified it as non-orientability of the real Bloch Hamiltonian \cite{Ahn2019}. This has to do with the property of a real Bloch Hamiltonian under translation by a reciprocal lattice vector $\G$. If we write $H(\k+\G)=O_\G H(\k) O_\G^{-1}$, with $O_\G$ an orthogonal matrix, then if $\det O_\G=-1$ the model is said to be non-orientable (and orientable otherwise). For the Lieb-kagom\'e model in basis II and for $\G=2\pi \mathbf{u}_x$, one has
\beq
H_{II}(\k+2\pi \mathbf{u}_x)&=& \left(
  \begin{array}{ccc}
   0 & -2\cos\frac{k_x}{2} & -2 t \cos\frac{k_x+k_y}{2}) \\
    \ldots & 0 & 2\cos\frac{k_y}{2} \\
   \ldots & \ldots & 0 \\
  \end{array}
\right)\nonumber \\
&=&O_x H_{II}(\k) O_x^{-1},
\eeq
with the orthogonal matrix $O_x=\text{diag}(-1,1,1)$. The fact that $\det O_x = -1$ means that it is non-orientable. This is also true of the translation by the reciprocal lattice vector $2\pi \mathbf{u}_y$:
\beq
H_{II}(\k+2\pi \mathbf{u}_y) = O_y H_{II}(\k) O_y^{-1},
\eeq
with the orthogonal matrix $O_y=\text{diag}(1,1,-1)$ such that $\det O_y= -1$. However for the translation by $\G=(2\pi,2\pi)$, we find that
\beq
H_{II}(\k+2\pi \mathbf{u}_x + 2\pi \mathbf{u}_y) = O_{x+y} H_{II}(\k) O_{x+y}^{-1},
\eeq
with the orthogonal matrix $O_{x+y}=\text{diag}(-1,1,-1)$ such that $\det O_{x+y}= 1$.

Such an analysis in terms of a real Bloch Hamiltonian does not apply to the case of the generalized Lieb-kagom\'e model of Appendix \ref{app:glk}. In this case, one has to use the winding vector concept.

\end{document}